\documentclass[useAMS,usenatbib]{mn2e}

\usepackage{graphicx,amssymb}
\usepackage[normalem]{ulem}
\usepackage[usenames,dvipsnames]{xcolor}
\usepackage{soul}

\newcommand{\rev}{}
\newcommand{\revv}{}
\newcommand{\revvv}{ }

\title[Migration in white dwarf discs]
{Planetesimals drifting through dusty and gaseous white dwarf debris discs: Types I, II and III-like migration}

\author[]{Dimitri Veras$^{1,2,3}$\thanks{E-mail: dimitri.veras@aya.yale.edu},
Shigeru Ida$^{4}$,
Evgeni Grishin$^{5,6}$,
Scott J. Kenyon$^{7}$,
\newauthor
Benjamin C. Bromley$^{8}$
\\
$^{1}$Centre for Exoplanets and Habitability, University of Warwick, Coventry CV4 7AL, UK
\\
$^{2}$Centre for Space Domain Awareness, University of Warwick, Coventry CV4 7AL, UK
\\
$^{3}$Department of Physics, University of Warwick, Coventry CV4 7AL, UK
\\
$^{4}$Earth-Life Science Institute, Tokyo Institute of Technology, Meguro, Tokyo 152-8550, Japan
\\
$^{5}$School of Physics and Astronomy, Monash University, Clayton, VIC 3800, Australia
\\
$^{6}$OzGrav: Australian Research Council Centre of Excellence for Gravitational Wave Discovery, Clayton, VIC 3800, Australia
\\
$^{7}$Smithsonian Astrophysical Observatory, 60 Garden Street, Cambridge, MA 02138, USA
\\
$^{8}$Department of Physics and Astronomy, University of Utah, 201 JFB, Salt Lake City, UT 84112, USA
}

\begin{document}
\label{firstpage}
\pagerange{\pageref{firstpage}--\pageref{lastpage}}
\maketitle

\begin{abstract}
The suite of over 60 known planetary debris discs which orbit white dwarfs, along with detections of multiple minor planets in these systems, motivate investigations about the migration properties of planetesimals embedded within the discs. Here, we determine whether any of the migration regimes which are common in  {\revv(pre-)main-sequence} protoplanetary discs, debris discs and ring systems could be active and important in white dwarf discs. We investigate both dust-dominated and gas-dominated regions, and quantitatively demonstrate that Type I and Type II migration, as well as their particulate disc analogues, are {\rev too slow to be relevant} in white dwarf discs. However, we find that the analogue of Type III migration for particulate discs may be {\rev rapid} in the dusty regions of asteroid- or moon-generated ($>10^{18}$~kg) white dwarf discs, where a planetesimal exterior to its Roche radius may migrate across the entire disc within its lifetime. This result holds over a wide range of disc boundaries, both within and exterior to $1R_{\odot}$, and such that the probability of migration occurring increases with higher disc masses. 
\end{abstract}


\section{Introduction}

A hallmark of the planet formation process is the migration of pebbles, boulders and embryos within circumstellar discs. Even before the discovery of extrasolar planets, the physical mechanism of migration was explored for protoplanets orbiting the Sun \citep{linpap1986} and protosatellites orbiting the solar system planets \citep{goltre1980}. {\rev The migration process is now thought to have been pivotal in shaping the architecture of at least the outer solar system \citep{malhotra1993,tsietal2005}.}

More recently, migration studies have overwhelmingly focussed on exoplanet formation \citep[e.g.][]{klenel2012,baretal2016,paaetal2022,raymor2022}, especially given the striking images of structure-laden protoplanetary discs \citep[e.g.][]{andrews2020,benetal2022,pinetal2022}. {\rev Also, migration models represent crucial components of population synthesis investigations \citep{idalin2004,idalin2008,moretal2012a,moretal2012b,emsetal2021}.}

Now, a different class of discs has emerged for which migration processes may be relevant: discs which orbit white dwarfs and are composed of the remnants of planetary bodies. These debris discs are fundamentally different from protoplanetary discs in size, composition and formation channel.

\begin{table*}
 \centering
 \begin{minipage}{180mm}
   \centering
   \caption{{\revv Broad comparison between protoplanetary discs and compact white dwarf debris discs}. 
}
   \label{CompTable}
  \begin{tabular}{cccccc}
\hline
 {\rm Disc}  & {\rm Inner}  &  {\rm Outer}  & {\rm Total} & {\rm Dust/} & {\rm Height}         \\[2pt]
 {\rm type}  & {\rm radius} &  {\rm radius} & {\rm mass}  & {\rm gas}   & {\rm aspect ratio}   \\[2pt]
\hline
\hline

{\rm Protoplanetary discs}                                     &  $\sim 10^{-1}$~{\rm au}              & $\sim 10^2$~{\rm au}                    & $\lesssim 10^{-1} M_{\star}$   &  $10^{-2}$  &  $0.05$          \\[2pt]
{\rm Compact white dwarf debris discs} &  $\sim 10^{-4}-10^{-2}$~{\rm au}  & $\sim 10^{-3}-10^{-1}$~{\rm au}  & $\lesssim M_{\rm planet}$  &  $0-\infty$      &  {\rm unknown}   \\[2pt]

\hline
\end{tabular}
\end{minipage}
\end{table*}


{\rev To understand the origin of these discs, consider} a planetary system which emerges from the giant branch phases of stellar evolution: any surviving planetary remnants must reside at a distance of at least a few au to have avoided tidal engulfment \citep{kunetal2011,musvil2012,norspi2013,viletal2014,madetal2016,ronetal2020}. In a few cases, these remnants, or their direct precursors, have been observed \citep{suetal2007,bonetal2013,bonetal2014,lovetal2022,maretal2022} despite the observational challenges of doing so. The surviving material may comprise the entire size spectrum of planets down to dust, and may be highly dispersed over the range 1-100 au due to radiative forces from the giant branch star \citep{veretal2014b,veretal2015a,veretal2019a,versch2020,feretal2022}.

Some of this planetary material then definitely makes {\revv its} way to the white dwarf, as {\revv currently} seen by the metal ``pollution" present in the photospheres {\revv of 15-50 per cent of white dwarfs} \citep{zucetal2010,koeetal2014,obretal2023}. {\revv The actual fraction of white dwarfs which are polluted at some point in their lives is likely higher, potentially up to 100 per cent.} The actual mechanisms for this inward transport are varied and unlikely to be mutually exclusive; see \cite{veras2021} for a review and \cite{broetal2022} for a recent canonical picture. Before this inward transport takes place, the debris structures on au-scales can persist for any amount of time depending on the physical details of these discs \citep{verhen2020}.

Upon reaching the close vicinity of the white dwarf, planetary material dissociates in a variety of ways, each of which forms its own type of debris disc. One type of disc, which is purely gas and resides on $10^{-2}-10^{-1}$ au scales, forms from the evaporation of planetary atmospheres around very hot white dwarfs \citep{ganetal2019,schetal2019}. Another type of disc, which is primarily dust and resides on $10^{-2}$ au scales, results from rotational fission of minor bodies on highly eccentric orbits \citep{makver2019,veretal2020}. The type of disc which is most easily and readily observed, on scales of $10^{-3}-10^{-2}$~au, results from breakup at the Roche sphere of the white dwarf \citep{graetal1990,jura2003,debetal2012,veretal2014,veretal2015b,malper2020a,malper2020b,lietal2021}\footnote{{\rev Another type of disc which may form around a white dwarf -- but is not composed of planetary remnants -- may be generated from the winds of a companion star \citep{perets2011,perken2013}, but only if the stellar binary is sufficiently close \citep{debes2006,veretal2018}. Hence, for single white dwarfs or binaries with separations greater than a few au, the debris disc can be assumed to be generated from planetary material.}}.

{\rev 
Over 60 white dwarf debris discs of this last type have already been observed as a result of infrared excess emission on spectral energy distributions \citep[e.g.][]{zucbec1987,graetal1990,becetal2005,baretal2014,rocetal2015,farihi2016,wiletal2019,swaetal2020}, which is an indication of the presence of dust. Of these discs, over 20 also have a detectable gaseous component \citep[e.g.][]{ganetal2006,denetal2018,ganetal2019,denetal2020,manetal2020,manetal2021,meletal2020,genetal2021}, although sublimation at the inner rim is a likely feature of every one of these discs {\revv \citep{rafikov2011a,rafikov2011b,metetal2012,steetal2021,okuetal2023}}. Gas which is present elsewhere in the disc may instead primarily form from collisions, as opposed to sublimation \citep{faretal2018b,swaetal2021}.
}

Investigations which have focussed on disc evolution, as opposed to formation, have featured processes which circularize the disc or keep it eccentric \citep{ocolai2020,malamudetal2021,treetal2021} and the effects of replenishment \citep{kenbro2017a,kenbro2017b,griver2019} and erosion \citep{rozetal2021}, but rarely migration of bodies through the disc. One of the most extensive studies on migration through a white dwarf disc was by \cite{ocolai2020}, who primarily focussed on tidal migration of planetesimals on highly eccentric orbits. As opposed to discs around main-sequence stars {\revv (see Table \ref{CompTable} for a comparison)}, large objects in $10^{-3}-10^{-2}$~au-scale white dwarf debris discs might be subject to {\rev torques} with the star, which could represent the primary driver of migration \citep{veretal2019b,verful2019} as opposed to interactions with the disc itself.

Nevertheless, understanding if planetesimal-disc interactions could instigate migration within white dwarf discs is important, and has been rarely addressed explicitly. 

Observationally, over half a dozen white dwarfs showcase photometric transit signatures or emission line shifts which are indicative of the presence of minor planets and their fragments \citep{vanetal2015,manetal2019,guietal2021,vanetal2020,vanetal2021,faretal2022}. {\revv They indicate that small bodies which are detected or inferred to orbit white dwarfs commonly co-exist with the dust and gas in the discs. Furthermore, the observations of WD 1054-226 \citep{faretal2022} have been interpreted as showing a resonant interaction between a minor planet and a dust disc; these resonant interactions are a common outcome of migration. Additionally} \cite{vanetal2018} demonstrated how {\revv resonant interactions can occur with} planetesimals that newly form, in a second-generation fashion, within a white dwarf disc which is viscously spreading outward to allow for coagulation beyond the Roche radius.

{\rev
For discs around main-sequence stars, the total disc mass has proven to be a key component in determining migration regimes and timescales. The same will be true for discs around white dwarfs, except that observational mass constraints for these discs are much poorer. The most robust constraints are obtained from the known systems with photometric transit signatures, otherwise measurements of accretion rates or convection zone masses would need to be coupled with assumptions about the highly uncertain disc lifetime. In systems with transiting debris, one may compute of the mass of the transiting cloud \citep{xuetal2016,vanetal2020,vanetal2021}, through which a lower limit to the disc mass may be obtained ($\sim 10^{15}-10^{16}$~kg). Fragments in the debris and the shape of the transit curve provide constrains on the mass of the progenitor body \citep{rapetal2016,veretal2017,guretal2017,duvetal2020}, which in one case is $10^{19}-10^{21}$~kg. In principle, the disc mass can be any higher value {\revv up to that which is set by the Toomre stability criterion \citep[Fig. 2 of][]{vanetal2018}}.  
}

\begin{figure}
\includegraphics[width=8cm]{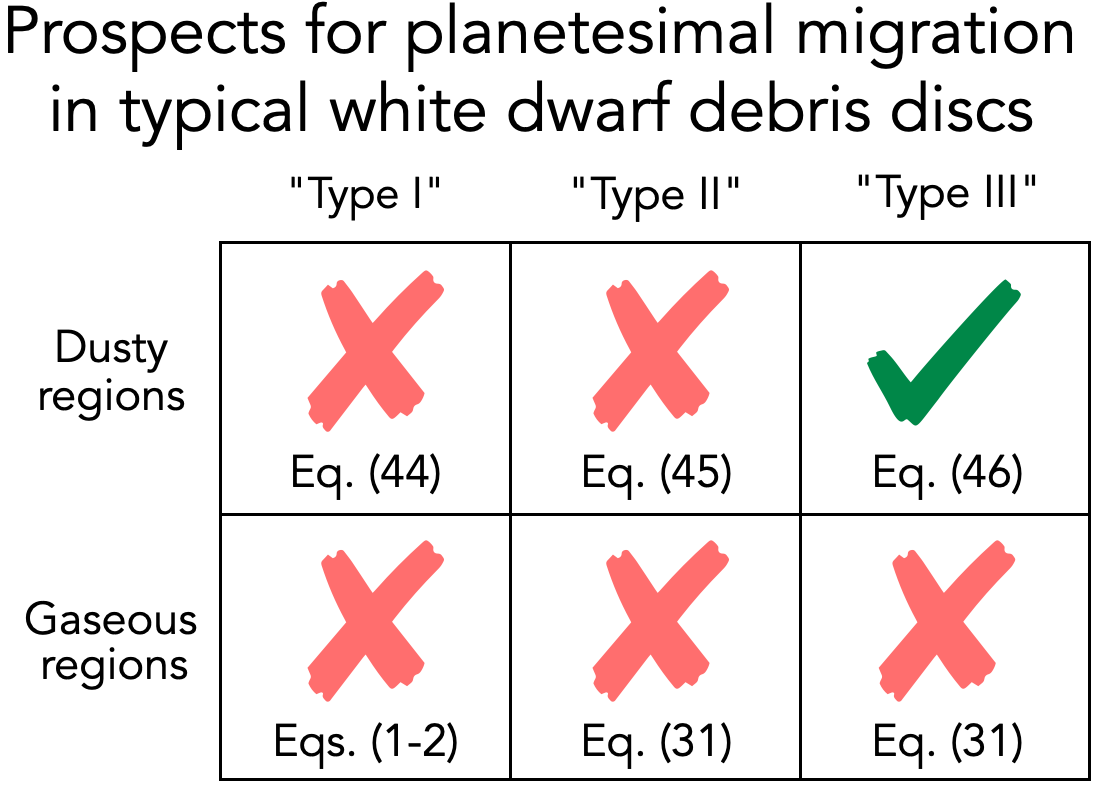}
\caption{
Our main result. Shown are the migration regimes for which planetesimals drift non-negligibly within white dwarf debris discs, which may contain both dust-dominated and gas-dominated regions. This chart illustrates that only corotation-induced runaway Type III migration in particulate (dust-dominated) disc regions {\rev is relevant}. 
}
\label{Tab1}
\end{figure}

Here, we aim to perform a general migration study covering all possible regimes in order to address common inquiries about the feasibility of Type I, II and III migration within white dwarf debris discs. We actually start by presenting the conclusion of our study now in Fig. \ref{Tab1}. The remainder of the document is devoted to supporting the claims in that figure. First, we need to define some terminology and physics (Section 2). Then, we will explore gas-dominated (Section 3) and dust-dominated (Section 4) regimes separately. We will discuss our results in Section 5 and conclude in Section 6.

\section{Migration terminology and regimes}

As already indicated, the migration of planets within protoplanetary and main-sequence debris discs has a long and rich history. As a result, well-established migration regimes used for these discs are often used as points of reference, as we do here.

Over time, that field has coalesced around three terms for describing different migration regimes in gas-rich (protoplanetary) discs: ``Type I", ``Type II" and ``Type III". Equivalent migration regimes in dust-rich (debris) discs, or more generally particulate discs, do not have names. Regardless, for the purposes of this paper, we will use Type I, Type II and Type III for both gas- and dust-dominated regions, for both ease of use and because the physics behind these regimes is the same.

All three types refer solely to disc-planet interactions generating migration of the planet through torques. In Type I migration {\rev \citep{ward1997,tanetal2002}}, the planet is not large enough to {\revv open a gap} and nearly always drifts inward. In Type II migration {\rev \citep{linpap1993,bryetal1999,kley1999,lubetal1999}}, the planet is large enough to change the disc structure by creating a gap, after which the planet and the gap migrate usually inwards, but sometimes outwards. Type III migration {\rev \citep{maspap2003,artymowicz2004,pepetal2008a,pepetal2008b,pepetal2008c}} is an accelerated intermediate version between Types I and II {\rev where the corotation torque is paramount. This migration type} features a planet large enough to open a partial gap, and allows for either inwards or outwards migration, but at an accelerated rate due to local disc conditions.

Which migration regime is applicable depends on the mass of the planet, mass of the star, location of the planet in the disc, and the disc properties, including its mass, geometry, and physical aspects of its constituents. These dependences may be also applied to white dwarf discs, with the caveat that the properties of most of these discs remain largely unknown, and common ways of parametrizing discs around main-sequence stars may not be realistic for some white dwarf discs. Further, migrating objects in white dwarf discs may better be described as planetesimals rather than planets, and we will do so henceforth.

For both gas-dominated and dust-dominated discs, the applicable migration regime is determined by (i) the size ratio of the local disc scale height to the Hill radius of the migrating planetesimal, and (ii) the depth of the gap formed (if any), which is dependent on disc viscosity. This latter point is one area where the gas-dominated and dust-dominated cases differ, because viscosity is defined differently for the different phases of matter.

Another aspect which differs between the gas- and dust-dominated cases is the migration rate itself. Over the last three decades analytical prescriptions for these rates have been continuously refined, particularly with the advent of detailed numerical simulations from which empirical relations can be extracted. {\rev Extra physics has also been added, helping to differentiate the regimes further. For example, gas dynamical friction has been shown to rapidly migrate large planetesimals \citep{mutetal2011,griper2015,griper2016}, and further motivates a determination of whether any quick migration regimes in white dwarf discs are possible.}

For this investigation, we sought sufficiently general analytical migration prescriptions that would be applicable across all these regimes and could be treated in a uniform manner. For gas-dominated discs, \cite{idaetal2020} provided a {\rev widely applicable Type I prescription which is valid for both sub-sonic and super-sonic migration, as well as for isothermal and non-isothermal discs; for the Type II and III cases, the gap-opening criteria of \cite{kanetal2018} and \cite{idaetal2018} sufficiently addresses both cases for our purposes}. For dust-dominated discs, \cite{broken2013} provided migration rates for each regime of motion {\rev (Type I, Type II, Type III)} which may be solved explicitly for the semi-major axis evolution of the planetesimal, {\rev independent of the temperature profile of the disc}.

The formalism in these studies will be used as the basis for the equations we adapt to white dwarf debris discs, and we perform this adaption in different ways. In Section 3, for the \cite{idaetal2020} formalism, we will start with the migration rate timescales and proceed to show that they are unfeasible for white dwarf discs. Alternatively, in Section 4, we start with a set of input variables and show how they lead to the migration rates from \cite{broken2013}, one of which is applicable for these discs.

\section{Gas-dominated discs}

In this section we describe migration within gas-dominated parts of white dwarf debris discs. Our thesis is that migration is {\rev negligible in all cases} in this context. 

\subsection{Type I migration: complete expressions}

In order to prove this theory, we begin by writing down the semimajor axis decay timescales $\tau_{a}(r_0)$ for a planetesimal in a disc from \cite{idaetal2020}. {\rev Importantly, these timescales are applicable for both the sub-sonic and super-sonic migration regimes.} For {\rev Type I migration in isothermal discs, the decay timescale is} 

\begin{equation}
\tau_{a}(r_0) = \frac{t_{\rm wave}(r_0)}{C_{\rm T} h(r_0)^2}  
 \left(1 + \frac{C_{\rm T}}{C_{\rm M}} \sqrt{\hat{e}(r_0)^2+ \hat{i}(r_0)^2} \right),
\label{TypeIandIIgas}
\end{equation}

\noindent{}and {\rev the decay timescale for Type I migration in non-isothermal discs is}

\[
\tau_{a}(r_0) = \frac{t_{\rm wave}(r_0)}{2h(r_0)^2}
\]

\[
\ \ \ \ \ \ \ \ \ \times 
\bigg[
C_{\rm P} \left(1 + \frac{C_{\rm P}}{C_{\rm M}} \sqrt{\hat{e}(r_0)^2+ \hat{i}(r_0)^2} \right)^{-1}
\]

\begin{equation}
\ \ \ \ \ \ \ \ \ 
-\frac{\Gamma_{\rm C}}{\Gamma_0} {\rm exp}\left( -\frac{\sqrt{e(r_0)^2 + i(r_0)^2}}{e_{\rm cor}(r_0)} \right)
\bigg]^{-1},
\label{TypeIIIgas}
\end{equation}

\noindent{}where the density wave timescale

\begin{equation}
t_{\rm wave}(r_0) = \left(\frac{M_{\star}}{M_{\rm pl}}\right)
                                 \left(\frac{M_{\star}}{\Sigma(r_0)r_{0}^2}\right)
                                  \frac{h(r_0)^{4}}{\Omega_{\rm K}(r_0)}.
\end{equation}

In these equations, $r_0$ represents the initial separation of the planetesimal from the white dwarf and $M_{\rm pl}$ and $M_{\star}$ represent the mass of the planetesimal and white dwarf, respectively. The orbital angular frequency at the location of the planetesimal is

\begin{equation}
\Omega_{\rm K}(r_0) = \sqrt{\frac{\mathcal{G} M_{\star}}{r_{0}^3}}
,
\end{equation}

\noindent{}{\revv where $\mathcal{G}$ is the gravitational constant.}

The values $e(r_0)$ and $i(r_0)$ represent the planetesimal's initial eccentricity and inclination at its initial separation. Further, by defining $H(r_0)$ as the height of the disc at the planetesimal's initial separation, we can establish the disc height aspect ratio as

\begin{equation}
h(r_0) = \frac{H(r_0)}{r_0}
,
\end{equation}

\noindent{}the corresponding inclination aspect ratio as

\begin{equation}
\hat{i}(r_0) = \frac{i(r_0)}{h(r_0)},
\end{equation}

\noindent{}the corresponding eccentricity aspect ratio as

\begin{equation}
\hat{e}(r_0) = \frac{e(r_0)}{h(r_0)},
\end{equation}

\noindent{}and the eccentricity due to corotation as

\begin{equation}
e_{\rm cor}(r_0) = 0.01 + \frac{h(r_0)}{2}
.
\end{equation}

The $C$ coefficients are given by

\begin{equation}
C_{\rm P} = 2.5 - 0.1p + 1.7q,
\end{equation}

\begin{equation}
C_{\rm T} = 2.73 + 1.08p,
\end{equation}

\begin{equation}
C_{\rm M} = 6 \left(2p - q + 2\right)
,
\end{equation}

\noindent{}where $p$ and $q$ are power-law exponents which define the surface density and temperature profile of the disc as

\begin{equation}
\Sigma(r_0) = \Sigma_{\rm norm} \left( \frac{r_0}{R_{\odot}} \right)^{-p},
\label{SurfDen}
\end{equation}

\noindent{}and

\begin{equation}
T(r_0) = T_{\rm norm} \left( \frac{r_0}{R_{\odot}} \right)^{-q}.
\end{equation}

\noindent{}The surface density normalization constant is

\[
\Sigma_{\rm norm} = 
\frac{M_{\rm d}\left(2-p\right)}
        {2\pi R_{\odot}^p}
\left[  
r_{\rm out}^{2-p} - r_{\rm in}^{2-p}
\right]^{-1}, \ \ \ \ p \ne 2
\]

\begin{equation}
\ \ \ \ \ \ \ \ \,   = 
\frac{M_{\rm d}}
        {2\pi R_{\odot}^p}
\left[  
\ln{\left( \frac{r_{\rm out}}{r_{\rm in}} \right)}
\right]^{-1}, \ \ \ \ \ \ \ \ \ \ \ \   p = 2
\end{equation}

\noindent{}and an explicit expression for $T_{\rm norm}$ is not needed for our purposes. Also, $r_{\rm in}$ and $r_{\rm out}$ represent the inner and outer disc radius, and $M_{\rm d}$ represents the mass of the disc.



Finally, the ratio $\Gamma_{\rm C}/\Gamma_{0}$, which represents the total {\revv normalised} corotation torque and is present in equation (\ref{TypeIIIgas}), is given by

\begin{equation}
\frac{\Gamma_{\rm C}}{\Gamma_0} 
= 
\left( \frac{\Gamma_{\rm C, baro}}{\gamma \Gamma_0} 
+ \frac{\Gamma_{\rm C, ent}}{\gamma \Gamma_0}
\right) {\rm exp}\left[ -\frac{e(r_0)}{e_{\rm cor}(r_0)} \right]
.
\label{gam1}
\end{equation}

\noindent{}From \citep{braetal2018},

\[
\frac{\Gamma_{\rm C, baro}}{\Gamma_0} = \frac{1}{\gamma} 
                                          \left(\frac{3}{2} - p \right)
\]
\begin{equation}
\ \ \ \ \ \ \ \ \ \ \  \times \left[1.1 F(u_{\nu}) G(u_{\nu})
+      0.7 \left(1 - K(u_{\nu})\right) 
\right],
\label{gam2}
\end{equation}

\[
\frac{\Gamma_{\rm C, ent}}{\Gamma_0} = \frac{1}{\gamma} \left(\frac{3}{2} - \frac{7}{5}p  \right)
\]

\[
\ \ \ \ \ \ \ \ \ \ \times \bigg\lbrace 5.6 F(u_{\nu}) F(u_{\chi}) \sqrt{G(u_{\nu}) G(u_{\chi}) } 
\]

\begin{equation}
\ \ \ \ \ \ \ \ \ \ + 0.8 
 \sqrt{\left[1 -  K(u_{\nu}) \right] \left[1 - K(u_{\chi}) \right] } 
\bigg\rbrace,
\label{gam3}
\end{equation}

\noindent{}where $\gamma$ is the adiabatic exponent of the gas, $u_{\chi}$ is the dimensionless thermal diffusion saturation parameter, and $u_{\nu}$ is the dimensionless viscosity saturation parameter (see \citealt*{paaetal2011}, who instead refer to these variables as $p_{\chi}$ and $p_{\nu}$). $F$, $G$ and $K$ refer to the following auxiliary functions

\begin{equation}
F(u) = \frac{8I_{4/3}(u)}{3uI_{1/3}(u) + \frac{9}{2} u^2 I_{4/3}(u)}
\end{equation}

\noindent{}where $I$ is the modified Bessel function of the first kind,

\[
G(u) = \frac{16}{25} \left( \frac{45\pi}{8} \right)^{3/4} u^{3/2}, \ \ \ \ \ \ \ \ \ \ \ 
  u < \sqrt{\frac{8}{45\pi}}
\]

\begin{equation}
\ \ \ \ \ \ \ = 1-\frac{9}{25} \left( \frac{8}{45\pi} \right)^{4/3} u^{-8/3}, \ \ \ \ u \ge \sqrt{\frac{8}{45\pi}},
\end{equation}

\noindent{}and

\[
K(u) = \frac{16}{25} \left( \frac{45\pi}{28} \right)^{3/4} u^{3/2}, \ \ \ \ \ \ \ \ \ \ \ 
  u < \sqrt{\frac{28}{45\pi}}
\]

\begin{equation}
\ \ \ \ \ \ \ \ = 1-\frac{9}{25} \left( \frac{28}{45\pi} \right)^{4/3} u^{-8/3}, \ \ \ \ u \ge \sqrt{\frac{28}{45\pi}}.
\label{KBessel}
\end{equation}

\subsection{Type I migration: focus on density wave timescale}

Having written out the complete expressions for the semimajor axis decay timescale $\tau_a(r_0)$, we can now partition the analysis into two of its functional dependencies. {\rev The first dependency is the scaled density wave timescale $t_{\rm wave}(r_0)/h(r_0)^2$}, which is the subject of this subsection. The second dependency, which includes all other variables, is the subject of the next subsection.

Equations (\ref{TypeIandIIgas})-(\ref{TypeIIIgas}) reveal that $\tau_a(r_0) \propto t_{\rm wave}(r_0)/h(r_0)^2 $. Hence, let us re-express this timescale with different numerical values for different types of white dwarf debris discs. For ease of approximation, let us first assume $r_0 = (r_{\rm in} + r_{\rm out})/2$. Then

\[
t_{\rm wave}(r_0) = \left(\frac{M_{\star}}{M_{\rm pl}}\right)
                                 \left(\frac{M_{\star}}{\Sigma(r_0)r_{0}^2}\right)
                                  \frac{h(r_0)^{4}}{\Omega_{\rm K}(r_0)}
\]

\[
= \frac{2\pi {\mathcal G}^{-\frac{1}{2}}}{2-p}
          \left(r_{\rm out}^{2-p} - r_{\rm in}^{2-p} \right)
          \left[ \frac{r_{\rm in} + r_{\rm out}}{2} \right]^{p-\frac{1}{2}}
          \frac{h^4 M_{\star}^{\frac{3}{2}}}{M_{\rm pl} M_{\rm d}}
, \ \ \ \ \ p \ne 2
\]

\[
= 2\pi \mathcal{G}^{-\frac{1}{2}}
          \ln{\left( \frac{r_{\rm out}}{r_{\rm in}} \right)}
          \left[ \frac{r_{\rm in} + r_{\rm out}}{2} \right]^{\frac{3}{2}}
          \frac{h^4 M_{\star}^{\frac{3}{2}}}{M_{\rm pl} M_{\rm d}}
, \ \ \ \ \ \ \ \ \ \ \ \ \ \,  p = 2.
\]
\begin{equation}
\end{equation}

\begin{figure*}
\centerline{\Huge \underline{Schematic of pre-defined discs}}
\centerline{}
\centerline{
\includegraphics[width=17.0cm]{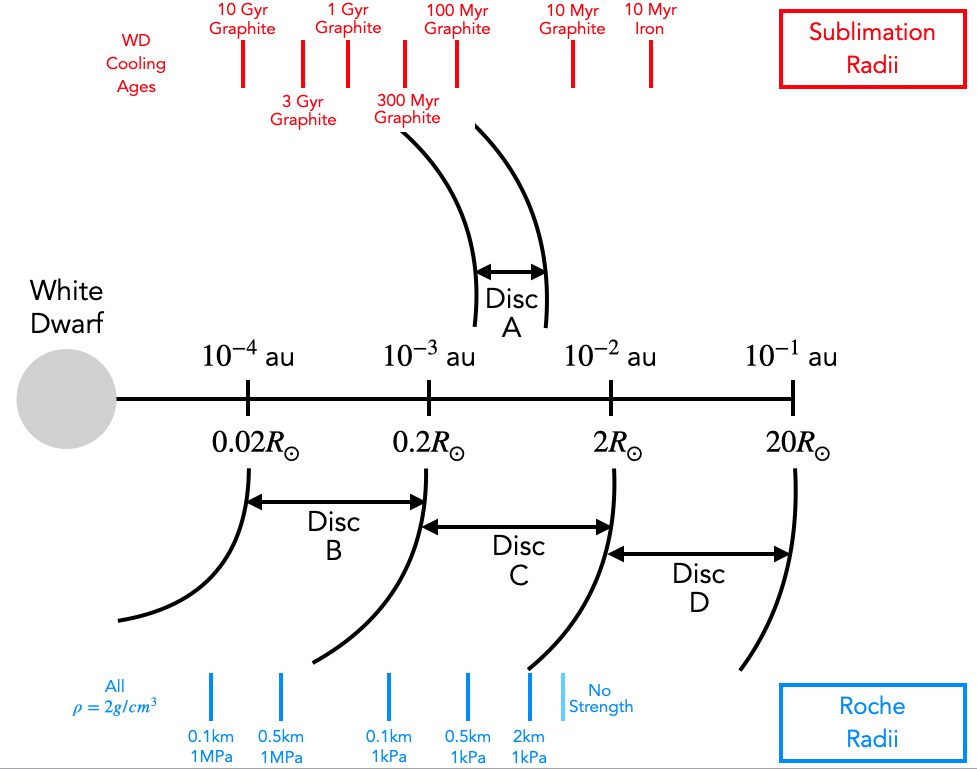}
}
\caption{
Cartoon of the radial extents of Discs A, B, C and D, which are used as fiducial discs throughout this paper. Also included are representative sublimation radii (on the top, in red, equation \ref{rsub}) and Roche radii (on the bottom, in blue, equation \ref{rroche}), where given values include planetesimal radii, cohesive strength and material composition, as well as white dwarf cooling age. {\revvv For perspective, an internal strength of 1MPa is comparable to that of ice, whereas a strength of 1kPa exceeds that of most rubble-pile asteroids.} These critical radii help determine locations at which migrating planetesimals can no longer survive the migration.
}
\label{DiscCartoon}
\end{figure*}

Now let us apply these expressions to four different types of discs, with the following representative radial extents {\rev (as already justified in Section 1;} see Fig. \ref{DiscCartoon}):

\begin{itemize}

\item Disc A: $2 \times 10^{-3} - 5 \times 10^{-3}$~au, or $\approx 0.43R_{\odot}-1.1R_{\odot}$, 

\item Disc B: $10^{-4} - 10^{-3}$~au, or $\approx 0.02R_{\odot}-0.2R_{\odot}$,

\item Disc C: $10^{-3} - 10^{-2}$~au, or $\approx 0.2R_{\odot}-2R_{\odot}$,

\item Disc D: $10^{-2} - 10^{-1}$~au, or $\approx 2R_{\odot}-20R_{\odot}$.

\end{itemize}

\noindent{}We also consider a range of surface density steepness profiles: in the following re-expressions of $t_{\rm wave}(r_0)$, we give timescale ranges for $p=0-5$ {\rev because observationally this parameter is unconstrained}. The resulting representative timescales become\footnote{{\revv Although Discs A and C overlap, their migration timescales are significantly different. The reason is that here the fiducial parameters are given in terms of the total disc mass rather than surface density.}}

\[
\frac{t_{\rm wave}(r_0)}{h(r_0)^2}^{\rm (Disc \, A)} \approx 42-81~{\rm Gyr} 
\left( \frac{h}{10^{-3}} \right)^2
\left( \frac{M_{\rm pl}}{10^{-2} M_{\rm Ceres}} \right)^{-1}
\]
\begin{equation}
\ \ \ \ \ \ \ \ \ \ \ \ \ \ \ \ \ \ \ \ \ \times \left( \frac{M_{\rm d}}{M_{\rm Ceres}} \right)^{-1}
\left( \frac{M_{\star}}{0.65M_{\odot}} \right)^{3/2}
\label{DiscADen}
\end{equation}

\[
\frac{t_{\rm wave}(r_0)}{h(r_0)^2}^{\rm (Disc \, B)} \approx 5-168~{\rm Gyr} 
\left( \frac{h}{10^{-3}} \right)^2
\left( \frac{M_{\rm pl}}{10^{-2} M_{\rm Ceres}} \right)^{-1}
\]
\begin{equation}
\ \ \ \ \ \ \ \ \ \ \ \ \ \ \ \ \ \ \ \ \ \times \left( \frac{M_{\rm d}}{M_{\rm Ceres}} \right)^{-1}
\left( \frac{M_{\star}}{0.65M_{\odot}} \right)^{3/2}
\end{equation}

\[
\frac{t_{\rm wave}(r_0)}{h(r_0)^2}^{\rm (Disc \, C)} \approx 160-5300~{\rm Gyr} 
\left( \frac{h}{10^{-3}} \right)^2
\left( \frac{M_{\rm pl}}{10^{-2} M_{\rm Ceres}} \right)^{-1}
\]
\begin{equation}
\ \ \ \ \ \ \ \ \ \ \ \ \ \ \ \ \ \ \ \ \ \times \left( \frac{M_{\rm d}}{M_{\rm Ceres}} \right)^{-1}
\left( \frac{M_{\star}}{0.65M_{\odot}} \right)^{3/2}
\end{equation}

\[
\frac{t_{\rm wave}(r_0)}{h(r_0)^2}^{\rm (Disc \, D)} \sim 10^{3}-10^{5}~{\rm Gyr}
\left( \frac{h}{10^{-3}} \right)^2
\left( \frac{M_{\rm pl}}{10^{-2} M_{\rm Ceres}} \right)^{-1}
\]
\begin{equation}
\ \ \ \ \ \ \ \ \ \ \ \ \ \ \ \ \ \ \ \ \ \times \left( \frac{M_{\rm d}}{M_{\rm Ceres}} \right)^{-1}
\left( \frac{M_{\star}}{0.65M_{\odot}} \right)^{3/2}
\label{DiscDDen}
\end{equation}

Hence, with the fiducial parameters that we chose, the timescales given in equations (\ref{DiscADen} - \ref{DiscDDen}) {\rev nearly all} exceed a Hubble time. However, before discarding migration prospects here as irrelevant, let us explore how changing these fiducial parameters change these timescales. 

\begin{enumerate}

\item First, the possible variation in the value of $M_{\star}$ can be safely ignored. The mass distribution of single white dwarfs sharply peaks at about $0.60-0.65M_{\odot}$ \citep{treetal2016,mccetal2020}. 

\item Second, the choice of $M_{\rm Ceres}$ {\revv ($\approx 9.1 \times 10^{20}~\ {\rm kg} \sim 10^{-4} M_{\oplus}$)} for the disc mass is reasonable given both observational and theoretical evidence. The mass of the progenitor asteroid in the WD~1145+017 system has been constrained to within an order of magnitude of $10^{20}$ kg \citep{rapetal2016,veretal2017,guretal2017}, although more sophisticated models of the asteroid structure may alter this value \citep{duvetal2020}.

There is no lower limit to reasonable values of $M_{\rm d}$, except that the lower the value, the longer the migration timescale. Mass values much higher than $M_{\rm Ceres}$, in the planet range, are unlikely but possible because of the ubiquity of white dwarf pollution, the relatively small number of planet-sized bodies in each system, and the rarity of events which can create a collision between a planet and a white dwarf \citep{veras2016,maletal2020a,maletal2020b,maletal2021,maletal2022}. Many moons, however, are larger than Ceres and may be gravitationally perturbed into the white dwarf \citep{payetal2016,payetal2017,trietal2022}. The mass of Io, for example, is two orders of magnitude higher than that of Ceres. However, equations (\ref{DiscADen}) - (\ref{DiscDDen}) illustrate that even a two order of magnitude decrease in the damping timescale is highly unlikely to generate non-negligible migration. 

\item Third, the choice of $10^{-2} M_{\rm Ceres}$ for the planetesimal mass is one of many acceptable choices. Because the breakup of the asteroid in WD~1145+017 appears to be messy and prolonged \citep{vanetal2015,ganetal2016,garetal2017,kjuetal2017,faretal2018a,izqetal2018,rapetal2018,xuetal2019}, we expect a wide mass spectrum of fragments. Further, that system also contains a dusty and gaseous disc around the same location of the debris, at around $1R_{\odot}$ \citep{cauetal2018,foretal2020}. Hence, the fragments are likely to be embedded in the disc.

Because $M_{\rm pl} < M_{\rm d}$, a substantial decrease in the timescales from equations (\ref{DiscADen}) - (\ref{DiscDDen}) can be achieved only if both $M_{\rm d}$ and $M_{\rm pl}$ increase. If, for example, $M_{\rm d} = M_{\rm Io}$ and $M_{\rm pl} = M_{\rm Ceres}$, then the decay timescale would decrease by four orders of magnitude, which is {\revv still longer than the lifetimes of such compact discs} \citep{giretal2012,verhen2020}. 

\item Fourth, the density wave timescale is most strongly dependent on $h$, which represents an unknown parameter in white dwarf debris discs.  

{\rev Regarding this last point, a longstanding assumption \citep{jura2003} is that white dwarf debris discs are thin due to dynamical settling, such that $h \sim 10^{-3}$ \citep{metetal2012}. However,} \cite{baletal2022} recently constrained the scale height of the G29-38 white dwarf debris disc {\rev at its inner edge} to be $H(r_{\rm in}) \ge 2.2 R_{\rm WD}$, where $r_{\rm in} = 96 R_{\rm WD}$, giving $h(r_{\rm in}) = 2 \times 10^{-2}$. {\rev This scale height constraint is for one location in the disc only; at other locations, this height is likely to be larger, even if, as the authors assumed, the disc is not flared.}

{\rev Regardless, for equations (\ref{DiscADen}) - (\ref{DiscDDen}), in an effort to be conservative, we chose a fiducial value of $h=10^{-3}$. The form of the equations then reveals that the disc would need to be even much flatter, by several orders of magnitude, in order to} make migration a realistic prospect. {\rev Recent theoretical developments would cast doubt on this possibility because of the likely prominence of disc particle collisions \citep{swaetal2021} and the short timescale of this process relative to dynamical settling timescales \citep{kenbro2017a}. However, before dismissing migration in this regime}, we first must consider the other dependencies of $t_{\rm wave}(r_0)$.

\end{enumerate}

\subsection{Type I migration: neglecting other contributions}

The last subsection demonstrated that in nearly all cases, the density wave timescale is longer than a Hubble time. However, the expressions for $\tau_a(r_0)$ (equations \ref{TypeIandIIgas}-\ref{TypeIIIgas}) are functions of many other variables. This subsection is devoted to determining how much these other variables can adjust the timescale set by $t_{\rm wave}(r_0)$.

\subsubsection{Representative values}

\begin{enumerate}

\item First, the aspect ratios $\hat{e}(r_0)$ and $\hat{i}(r_0)$ become large for thin discs, which would just increase the migration timescales for Type I migration. The largest values of these ratios would be on the order of unity. Also, the value of $e_{\rm cor}(r_0)$ lies in the approximate range $0.01-1$. 

\item Second, based on {\revv $p \in \left[0,5\right]$ and $q \in \left[0.50,0.75\right]$} {\rev \citep{kenhar1987,chigol1997,matetal2009}}, we obtain {\revv $C_{\rm P} \in \left[2.8,3.8\right]$, $C_{\rm T} \in \left[2.7,8.1\right]$ and $C_{\rm M} \in \left[9,69\right]$}.  

\item Third, {\rev for non-isothermal discs}, we can approximate the ratio $\Gamma_{\rm C}/\Gamma_0$ by adopting the range {\revv $u_{\nu} \in \left[0.1,10\right]$} and $u_{\chi} = \frac{3}{2} u_{\nu}$ \citep{paaetal2011}. We also assume {\revv $\gamma \in \left[7/5,5/3\right]$}, but given the functional dependence in equations (\ref{gam1})-(\ref{gam3}), we can immediately see that the choice of $\gamma$ within this range will have a negligible effect on the migration timescale. Hence, we set $\gamma=7/5$.

We perform the calculations with equations (\ref{gam1})-(\ref{KBessel}). For $p=0-5$, we obtain 

\begin{equation}
 \frac{\Gamma_{\rm C, baro}}{\gamma \Gamma_0} 
+ \frac{\Gamma_{\rm C, ent}}{\gamma \Gamma_0}
\in \left[-7.2, 2.2 \right],  \ \ \ \ \ \ \ \ \ \ \,  u_{\nu} = 0.1
\end{equation}

\begin{equation}
 \frac{\Gamma_{\rm C, baro}}{\gamma \Gamma_0} 
+ \frac{\Gamma_{\rm C, ent}}{\gamma \Gamma_0}
\in \left[-5.2, 1.6 \right],  \ \ \ \ \ \ \ \ \ \ \, u_{\nu} = 1.0
\end{equation}

\begin{equation}
 \frac{\Gamma_{\rm C, baro}}{\gamma \Gamma_0} 
+ \frac{\Gamma_{\rm C, ent}}{\gamma \Gamma_0}
\in \left[-0.035, 0.015 \right],  \ \ \ \ \  u_{\nu} = 10.0.
\end{equation}

\end{enumerate}

\subsubsection{Summary}

Overall, the results in Subsection 3.3.1 show that the timescale given by $t_{\rm wave}(r_0)$ for {\rev isothermal disc} migration cannot be decreased by the other variables in equations (\ref{TypeIandIIgas}) because the term in parenthesis is positive definite, given that $C_{\rm T} > 0$ and $C_{\rm M} > 0$. For {\rev non-isothermal disc} migration, the term in square brackets in equation (\ref{TypeIIIgas}) must be large (and positive) in order to decrease the migration timescale. However, the maximum value of this term is just max$(C_{\rm P})-$ min$(\Gamma_{\rm C}/\Gamma_{0})$exp($0$)$=11$, which is insufficient except in rare cases.

Hence, we conclude that {\it gas-driven Type I migration in white dwarf discs does not occur}. {\rev We have demonstrated this result mathematically. The primary physical reason, when compared to protoplanetary discs, is that discs which are present around white dwarfs are not massive enough to generate sufficiently high torques on an embedded planetesimal to force radial migration.} 

\subsection{Type I migration: Eccentricity and inclination damping timescales}

We have shown that {\rev Type I} radial migration in gas-rich white dwarf discs is negligible. However, a separate but related question is, how quickly do disc-planetesimal interactions damp the planetesimal's eccentricity and inclination? The answer {\revv depends on whether the planetesimal remains in the disc throughout the damping. In this case, the eccentricity and inclination damping timescales are} given by \citep{idaetal2020}

\begin{equation}
\tau_{e}(r_0) = 1.28 \, t_{\rm wave}(r_0) \left[1 + \frac{1}{15} \left(\hat{e}(r_0)^2+ \hat{i}(r_0)^2  \right)^{3/2} \right]
\label{EccDamp}
, 
\end{equation}

\begin{equation}
\tau_{i}(r_0) = 1.84 \, t_{\rm wave}(r_0) \left[1 + \frac{1}{21.5} \left(\hat{e}(r_0)^2+ \hat{i}(r_0)^2  \right)^{3/2} \right] 
\label{IncDamp}
.
\end{equation}

Both of these timescales operate in both the {\rev isothermal and non-isothermal} regimes. Because of the dependences on $t_{\rm wave}(r_0)$, $\hat{e}(r_0)$ and $\hat{i}(r_0)$, we can immediately see that white dwarf discs do not circularise planetesimal orbits or make them planar, at least from Type I disc-planetesimal torques alone (as opposed to gas drag or tides with the star, {\rev and especially from violent eccentricity damping immediately after a tidal disruption event; \citealt*{malamudetal2021}}). 

{\revv For planetesimals which do not remain embedded in the disc \citep{ocolai2020}, the eccentricity and inclination damping timescales will be much longer than the timescales given in equations (\ref{EccDamp})-(\ref{IncDamp}). For bodies on extremely eccentric orbits, such as those approaching a white dwarf disc for the first time, most of their time will be spent outside of the disc.}

\subsection{Types II and III migration}

{\rev

Having shown that Type I migration is ineffectual in gas-dominated regions of white dwarf discs, we now address Type II and Type III migration. The former requires a near-complete gap in the disc to be opened, while the latter requires a partial gap to be opened. In this short subsection, we demonstrate that white dwarf disc parameters do not allow for either situation to occur.

Both \cite{kanetal2018} and \cite{idaetal2018} computed a gap parameter, denoted $K(r_0)$, that allows one to determine whether or not a gap is formed, and its depth. Doing so requires us to define an $\alpha$ viscosity (which we have so far avoided). With this viscosity, the value of the gap parameter is 

\[
K(r_0) = \left(\frac{M_{\rm pl}}{M_{\star}}\right)^2
       \frac{1}{\alpha h(r_0)^{5}}
\]

\[
\ \ \ \ \ \ \ = 5 \times 10^{-6} 
          \left( \frac{M_{\rm pl}}{10^{-2} M_{\rm Ceres}} \right)^{2}
          \left( \frac{M_{\star}}{0.65M_{\odot}} \right)^{-2}
\]

\begin{equation}
\ \ \ \ \ \ \ \times \left( \frac{h(r_0)}{10^{-3}} \right)^{-5}
          \left( \frac{\alpha}{10^{-2}} \right)^{-1}.
\label{Kgap}
\end{equation}

The key condition to prevent a gap from opening is $K(r_0) \lesssim 25$ \citep{kanetal2018,idaetal2018}. Hence, equation (\ref{Kgap}) is sufficient alone to demonstrate that neither Type II nor Type III migration occur on important timescales; the physical conditions of a gas-dominated white dwarf disc do not allow for even a partial gap to open unless the disc is very thin. 

}





\section{Dust-dominated discs}

Now we consider regions of white dwarf debris discs which contain a negligible amount of gas. The dust-dominated regions may be considered as particulate discs. Because, in these discs, Type III migration is {\rev quick} (as we will show), and the conditions which prompt it need to be quantified, the development in this section will differ from that in Section 3. Here we will start afresh with a set of given quantities which will carry us through to the final result, using the development of \cite{verhen2020} and the results of \cite{broken2013}.


\subsection{Given variables}

We can treat the dust as a collection of discrete particles with radius $R_{\rm part}$ and density $\rho_{\rm part}$. The remainder of the disc is defined by {\revv six} variables: total mass, $M_{\rm d}$, inner radius $r_{\rm in}$, outer radius $r_{\rm out}$, {\revv power-law exponent for surface density, $p$,} root-mean-squared eccentricity of particles $\sigma_{e}$ and root-mean-squared inclination of particles $\sigma_{i}$. We assume that the planetesimal which is embedded in the disc has a mass of $M_{\rm pl}$ and semimajor axis of $a_{\rm pl}$, and is orbiting a star of mass $M_{\star}$. 

Our input choices of these {\revv 11} variables only will ultimately determine the migration regimes and rates of the planetesimal. 


\subsection{Intermediate quantities}

First, {\rev we will use these input quantities to} compute some intermediate quantities, primarily for the disc. A characteristic semimajor axis for the disc, $a_{\rm d}$, is

\begin{equation}
a_{\rm d} = \frac{1}{2}\left(r_{\rm in} + r_{\rm out} \right)
\end{equation}



\noindent{}and the radial velocity dispersion of the disc particles, $\sigma_{r}$, with

\begin{equation}
\sigma_{r} = \frac{\sigma_{e}a_{\rm d}\Omega_{\rm d}}{\sqrt{2}}
,
\end{equation}

\noindent{}where the effective orbital frequency of the disc $\Omega_{\rm d}$ is

\begin{equation}
\Omega_{\rm d} = \sqrt{\frac{\mathcal{G} M_{\star}}{a_{\rm d}^3}}.
\end{equation}

The disc's surface density, $\Sigma_{\rm d}$, {\revv is a function of distance from the star and follows the power-law prescription of equation (\ref{SurfDen})} as

\begin{equation}
\Sigma_{\rm d}(r) = \Sigma_{\rm norm} \left( \frac{r}{R_{\odot}} \right)^{-p}
\end{equation}

With the radial velocity dispersion of particles {\revv and surface density of the disc} in hand, we can now compute the disc's viscosity $\nu(r)$ with \citep{coofra1964,goltre1978,goltre1982} 

\begin{equation}
\nu(r) = \frac{\sigma_{r}^2 T_{\rm d}}{4\pi}
      \frac{\tau_{\rm d}(r)}{1 + \tau_{\rm d}(r)^2},
\end{equation}

\noindent{}where the optical depth in the disc, $\tau_{\rm d}(r)$, is

\begin{equation}
\tau_{\rm d}(r) = \frac{3\Sigma_{\rm d}(r)}{4 \rho_{\rm part} R_{\rm part}}
\end{equation}

\noindent{}and the characteristic orbital period of the disc $T_{\rm d}$ is

\begin{equation}
T_{\rm d} = \frac{2\pi}{\Omega_{\rm d}}.
\end{equation}

Finally, the orbital period of the planetesimal, $T_{\rm pl}$, is just

\begin{equation}
T_{\rm pl} = 2\pi\sqrt{\frac{a_{\rm pl}^3}{GM_{\star}}}.
\end{equation}

\subsection{Key scale lengths and regimes}

Having computed these intermediate quantities, we can now determine four key scale lengths as follows \citep{verhen2020,broken2013}

\begin{equation}
H = \frac{\sigma_{i} a_{\rm d}}{\sqrt{2}},
\end{equation}

\begin{equation}
r_{\rm Hill} = a_{\rm pl} \left( \frac{M_{\rm pl}}{3 M_{\star}} \right)^{1/3},
\end{equation}

\begin{equation}
r_{\rm gap}(r) = 0.4 \left(\nu(r) a_{\rm pl} T_{\rm pl} \right)^{1/3},
\end{equation}

\begin{equation}
r_{\rm fast}(r) = 1.7 a_{\rm pl}^2 \sqrt{\frac{\Sigma_{\rm d}(r)}{M_{\star}}}.
\end{equation}

As alluded to in Section 2, these four lengths are the ingredients that allow us to provide conditions on which migration regime is most applicable. The value $H$ is the vertical scale height of the particles that the planetesimal migrates through, $r_{\rm Hill}$ is the Hill radius of the planetesimal, {\revv $r_{\rm gap}(r)$ is the Hill radius ($r_{\rm Hill}$) of the smallest body that can open a gap in the particle disc}, and $r_{\rm fast}(r)$ is the maximum radius a planetesimal with Hill radius $r_{\rm Hill}$ can have in order to undergo Type III migration.

Although the physical transitions between the different regimes {\rev are} not necessarily straightforward \citep[see][for a more detailed explanation]{broken2013}, here we do not explore such details and seek to just identify the potential migration regimes in white dwarf debris discs. {\revv As a reminder, in particulate discs, the three migration types are similar to those in gaseous discs: Type I does not form a gap and is fast, Type II forms a large gap and is slow, and Type III forms a partial gap and is super-fast. Types I and II migration are driven by radial torques whereas Type III migration is driven by corotation torques.}

The conditions which determine the applicable migration regime are:

\begin{itemize}

\item If both $r_{\rm Hill} < H$ and $r_{\rm Hill} < r_{\rm gap}(r)$, then the migration regime is Type I.

\item If $r_{\rm Hill} > r_{\rm gap}(r)$, then the migration regime will be either Type II or Type III, with the latter applying only if, in addition, $r_{\rm Hill} > H$ and $r_{\rm Hill} < r_{\rm fast}(r)$.


\end{itemize}

\subsection{Migration rates}

The migration rate for each regime is as follows \citep{broken2013}:

\begin{equation}
\frac{da_{\rm I}(r)}{dt} = -\frac{64 \pi a_{\rm I}^2 \Sigma_{\rm d}(r) r_{\rm Hill}^3}
                              {M_{\star} H^2 T_{\rm pl}}
                      = -\frac{32 \mathcal{G}^{1/2} a_{\rm I}^{7/2} \Sigma_{\rm d}(r) M_{\rm pl}}
                              {3 H^2 M_{\star}^{3/2}}       
                              ,
\label{dadtMigI}
\end{equation}

\begin{equation}
\frac{da_{\rm II}(r)}{dt} = -\frac{16 \pi a_{\rm II}^2 \Sigma_{\rm d}(r) r_{\rm Hill}}
                              {M_{\star} T_{\rm pl}}
                       = -\frac{5.5 \mathcal{G}^{1/2} a_{\rm II}^{3/2} \Sigma_{\rm d}(r) M_{\rm pl}^{1/3}}
                              {M_{\star}^{5/6}}      
                              ,
\label{dadtMigII}
\end{equation}

\begin{equation}
\frac{da_{\rm III}(r)}{dt} = \pm \frac{5.3 \pi a_{\rm III}^3 \Sigma_{\rm d}(r)}
                             {M_{\star} T_{\rm pl}}
                        = \pm \frac{2.7 \mathcal{G}^{1/2} a_{\rm III}^{3/2} \Sigma_{\rm d}(r)}{M_{\star}^{1/2}} 
                         .
\label{dadtMigIII}
\end{equation}

{\revv Now we wish to} obtain an order-of-magnitude sense of the relevance of these different migration regimes. {\revv In order to do so} first we {\revv set $p=0$ and} re-express equations (\ref{dadtMigI})-(\ref{dadtMigIII}) as

\begin{equation}
\frac{da_{\rm I}}{dt} = -\frac{256 \mathcal{G}^{1/2} \left(r_{\rm out} + r_{\rm in} \right)^{3/2}}
                              {2^{7/2}\cdot 3\pi \left(r_{\rm out}^2 - r_{\rm in}^2\right)}
                        \left( \frac{a_{\rm I}}{a_{\rm d}} \right)^{7/2}
                        \left[
                        \frac{M_{\rm d} M_{\rm pl}}{M_{\star}^{3/2} \sigma_{i}^2}
                        \right],     
\label{dadtMigIV2}
\end{equation}

\begin{equation}
\frac{da_{\rm II}}{dt} = -\frac{5.5 \mathcal{G}^{1/2} \left(r_{\rm out} + r_{\rm in} \right)^{3/2}}
                              {2^{3/2}\pi \left(r_{\rm out}^2 - r_{\rm in}^2\right)}
                        \left( \frac{a_{\rm II}}{a_{\rm d}} \right)^{3/2}
                        \left[
                        \frac{M_{\rm d} M_{\rm pl}^{1/3}}{M_{\star}^{5/6}}
                        \right],     
\label{dadtMigIIV2}
\end{equation}

\begin{equation}
\frac{da_{\rm III}}{dt} = \pm\frac{2.7 \mathcal{G}^{1/2} \left(r_{\rm out} + r_{\rm in} \right)^{3/2}}
                              {2^{3/2}\pi \left(r_{\rm out}^2 - r_{\rm in}^2\right)}
                        \left( \frac{a_{\rm III}}{a_{\rm d}} \right)^{3/2}
                        \left[
                        \frac{M_{\rm d}}{M_{\star}^{1/2}}
                        \right].     
\label{dadtMigIIIV2}
\end{equation}

We can use the four types of discs that were defined just before equation (\ref{DiscADen}) in order to quantify these rates. The maximum migration rates are given by Disc B, which yields for Type I and Type II migration

\[
\frac{da_{\rm I}}{dt}^{({\rm Disc \, B})} = -5.1 \times 10^{-4} \frac{R_{\odot}}{{\rm Myr}}
                        \left( \frac{a_{\rm I}}{5 \times 10^{-3} \, {\rm au}} \right)^{\frac{7}{2}}
                        \left( \frac{\sigma_i}{10^{-3} \, {\rm rad}} \right)^{-2}
\]

\begin{equation}
\ \ \ \ \ \ \ \times \left( \frac{M_{\rm pl}}{10^{-2} M_{\rm Ceres}} \right)
       \left( \frac{M_{\rm d}}{M_{\rm Ceres}} \right)
       \left( \frac{M_{\star}}{0.65 M_{\odot}} \right)^{-\frac{3}{2}}
\label{dadtMigIV3}
\end{equation}

\noindent{}and

\[
\frac{da_{\rm II}}{dt}^{({\rm Disc \, B})} = -3.5 \times 10^{-3} \frac{R_{\odot}}{{\rm Myr}}
                        \left( \frac{a_{\rm II}}{5 \times 10^{-3} \, {\rm au}} \right)^{\frac{3}{2}}
\]

\begin{equation}
\ \ \ \ \ \ \ \times \left( \frac{M_{\rm pl}}{10^{-2} M_{\rm Ceres}} \right)^{\frac{1}{3}}
       \left( \frac{M_{\rm d}}{M_{\rm Ceres}} \right)
       \left( \frac{M_{\star}}{0.65 M_{\odot}} \right)^{-\frac{5}{6}}
.
\label{dadtMigIIV3}
\end{equation}

Both of these rates indicate that even if a white dwarf disc were to persist, and in a relatively steady state, for 1 Myr, then a planetesimal would migrate at most through about only 0.1 per cent of the disc. Such slow rates are effectively negligible given that compact debris discs are unlikely to survive for more than a Myr, unless very {\revv dynamically} cold \citep{giretal2012,verhen2020}. 

Hence, we focus on Type III migration, which has the potential to be much more consequential. In this regard, we compute representative migration rates for all four types of our fiducial discs, as follows

\[
\frac{da_{\rm III}}{dt}^{({\rm Disc \, A})} = \pm 6.7 \frac{R_{\odot}}{{\rm Myr}}
                        \left( \frac{a_{\rm III}}{3.5 \times 10^{-3} \, {\rm au}} \right)^{\frac{3}{2}}
\]

\begin{equation}
\ \ \ \ \ \ \ \ \ \ \ \ \ \ \ \ \ \ \ \ \times 
       \left( \frac{M_{\rm d}}{M_{\rm Ceres}} \right)
       \left( \frac{M_{\star}}{0.65 M_{\odot}} \right)^{-\frac{1}{2}}
,
\label{dadtMigIIIV3DiscA}
\end{equation}

\[
\frac{da_{\rm III}}{dt}^{({\rm Disc \, B})} = \pm 8.8 \frac{R_{\odot}}{{\rm Myr}}
                        \left( \frac{a_{\rm III}}{5 \times 10^{-3} \, {\rm au}} \right)^{\frac{3}{2}}
\]

\begin{equation}
\ \ \ \ \ \ \ \ \ \ \ \ \ \ \ \ \ \ \ \ \times 
       \left( \frac{M_{\rm d}}{M_{\rm Ceres}} \right)
       \left( \frac{M_{\star}}{0.65 M_{\odot}} \right)^{-\frac{1}{2}}
,
\label{dadtMigIIIV3DiscB}
\end{equation}

\[
\frac{da_{\rm III}}{dt}^{({\rm Disc \, C})} = \pm 2.8 \frac{R_{\odot}}{{\rm Myr}}
                        \left( \frac{a_{\rm III}}{5 \times 10^{-2} \, {\rm au}} \right)^{\frac{3}{2}}
\]

\begin{equation}
\ \ \ \ \ \ \ \ \ \ \ \ \ \ \ \ \ \ \ \ \times 
       \left( \frac{M_{\rm d}}{M_{\rm Ceres}} \right)
       \left( \frac{M_{\star}}{0.65 M_{\odot}} \right)^{-\frac{1}{2}}
,
\label{dadtMigIIIV3DiscC}
\end{equation}

\[
\frac{da_{\rm III}}{dt}^{({\rm Disc \, D})} = \pm 0.88 \frac{R_{\odot}}{{\rm Myr}}
                        \left( \frac{a_{\rm III}}{5 \times 10^{-1} \, {\rm au}} \right)^{\frac{3}{2}}
\]

\begin{equation}
\ \ \ \ \ \ \ \ \ \ \ \ \ \ \ \ \ \ \ \ \times 
       \left( \frac{M_{\rm d}}{M_{\rm Ceres}} \right)
       \left( \frac{M_{\star}}{0.65 M_{\odot}} \right)^{-\frac{1}{2}}
.
\label{dadtMigIIIV3DiscD}
\end{equation}

These rates showcase the potential for a planetesimal to migrate through a substantial fraction of a (dusty portion of) a disc over its lifetime, particularly if the disc mass is greater than that of Ceres. If the white dwarf disc was formed by the destruction of a moon such as Io, then the migration rates would be increased by two orders of magnitude.

Even such an increase, however, would yield a rate which is orders of magnitude slower than the rates \cite{broken2013} derived for planetesimal migration in Saturn's particulate rings. The reason is primarily because of the difference in mass of the central object in the system, despite the weak dependence of the Type III migration rate on the central mass.

\begin{figure*}
\centerline{\Huge \underline{Permissible planetesimal and disc masses}}
\centerline{}
\centerline{}
\centerline{
\includegraphics[width=8.5cm]{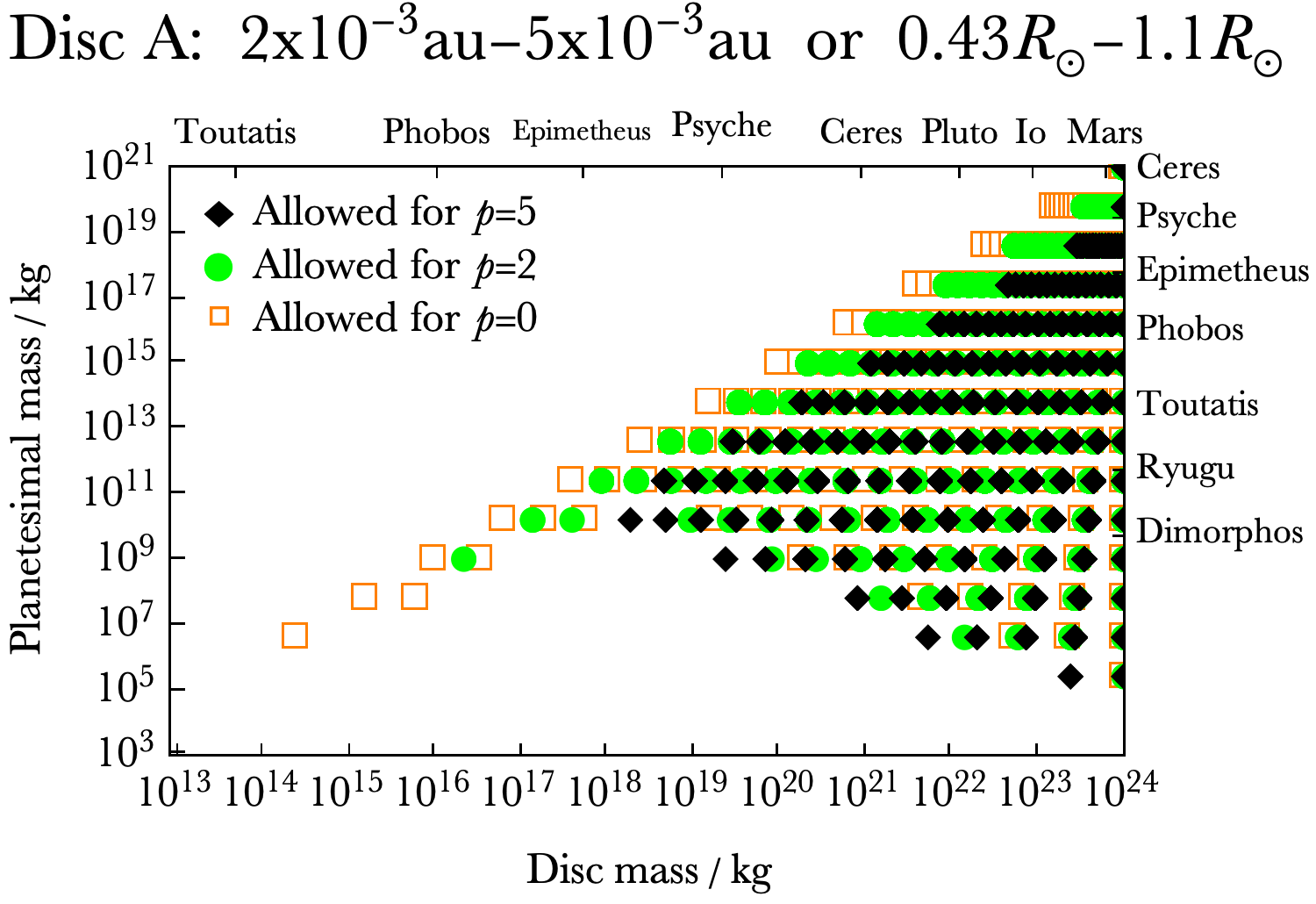}
\includegraphics[width=8.5cm]{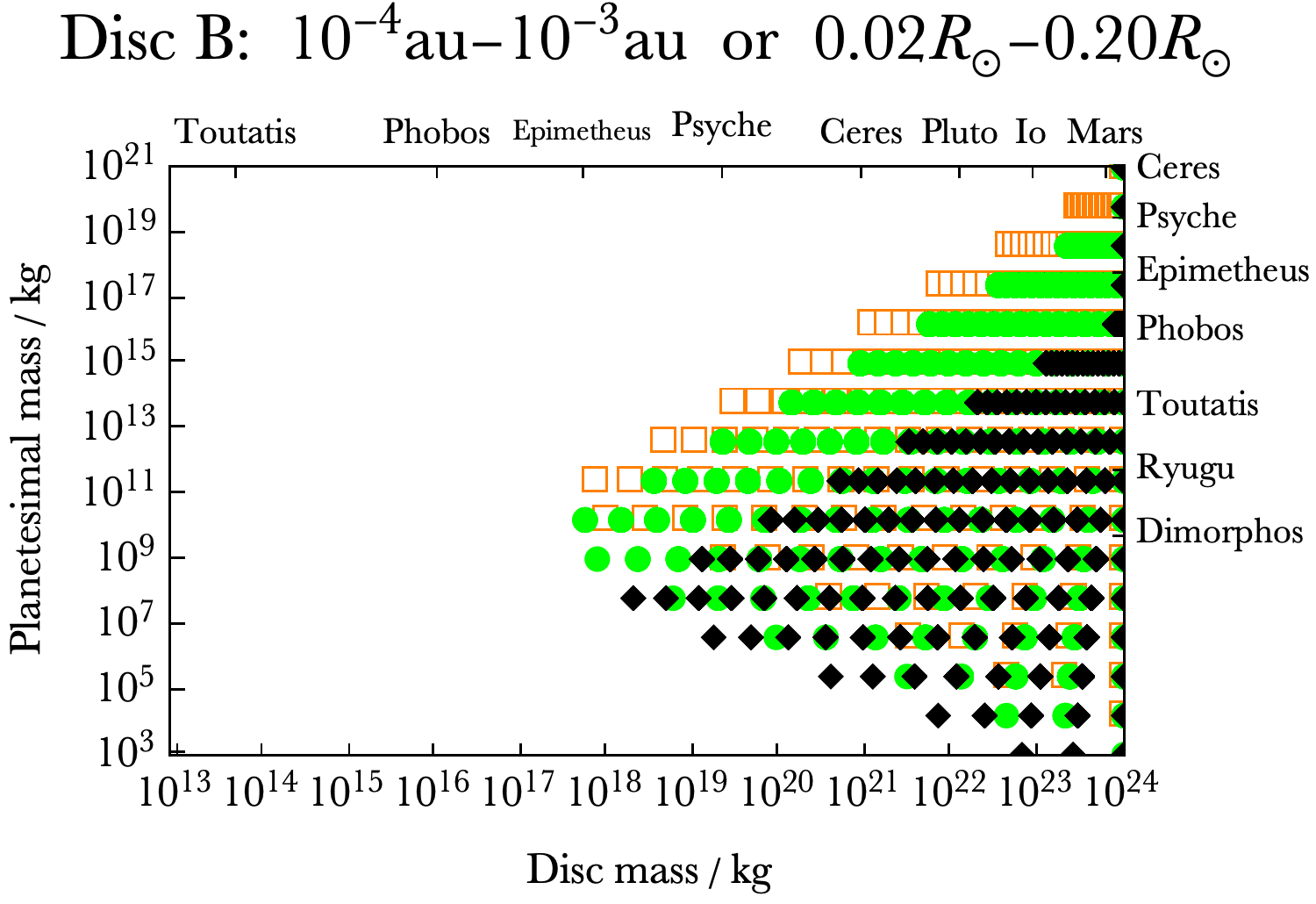}
}
\centerline{}
\centerline{
\includegraphics[width=8.5cm]{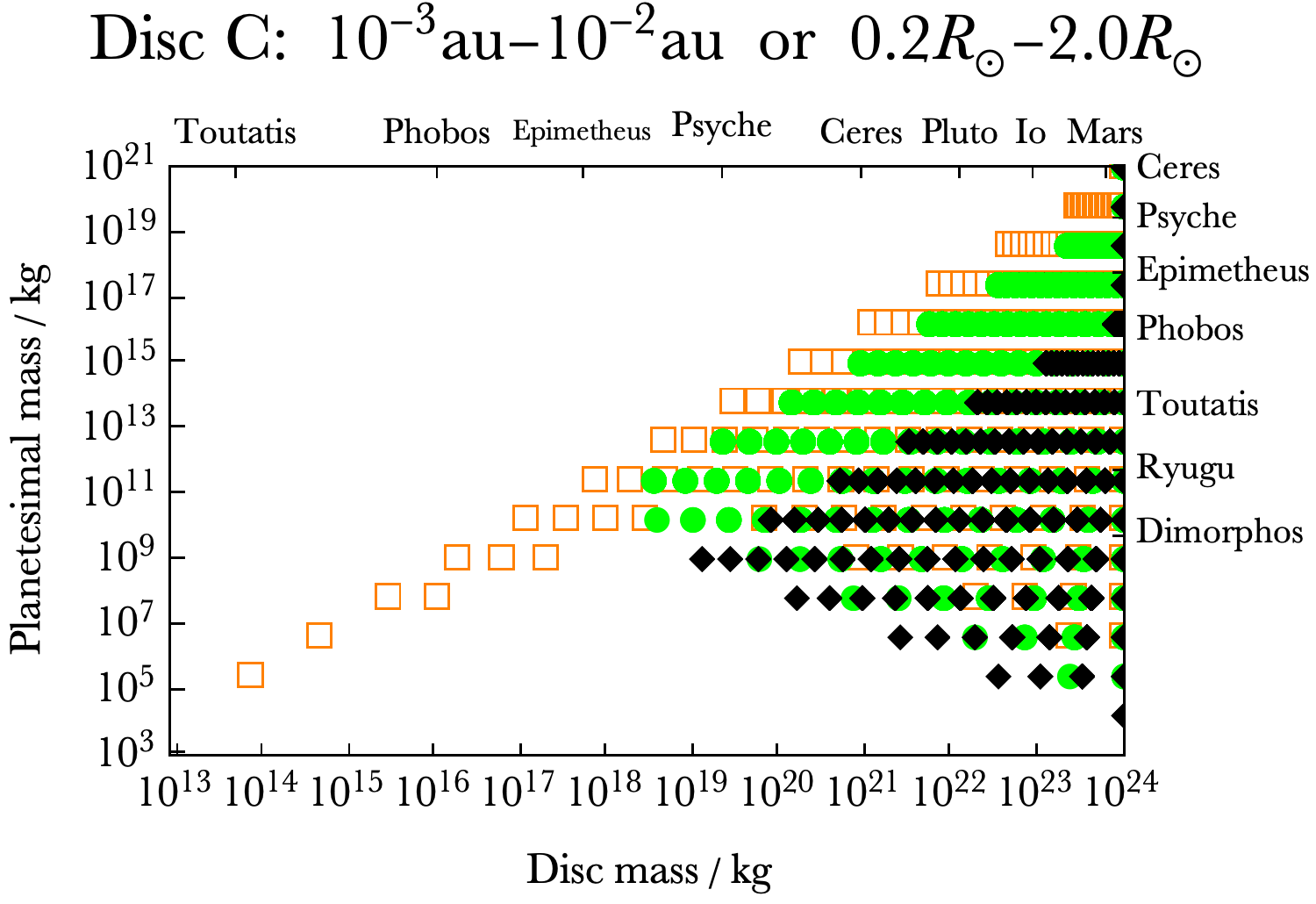}
\includegraphics[width=8.5cm]{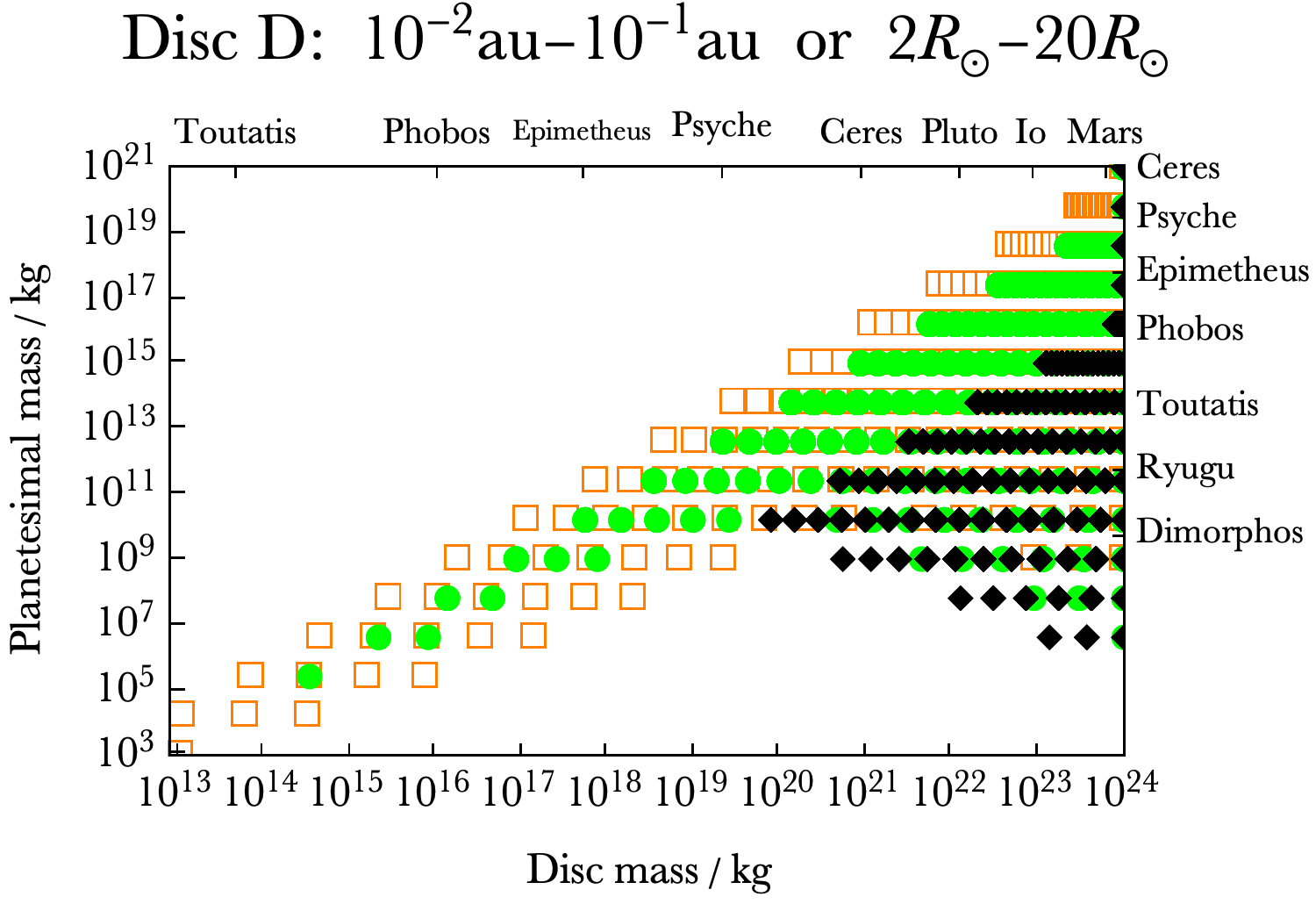}
}
\caption{
Combinations of planetesimal and disc masses (and power-law surface density exponents $p$) which allow for Type III migration to occur in dusty parts of white dwarf debris discs. Each symbol represents a permissible set {\rev of all {\revv 11} input variables; see Section 4.5}. The absence of a symbol indicates a region of parameter space that is not allowed. These plots illustrate that migration is effectively possible only in discs more massive than those created by planetesimals with Epimetheus-like masses. 
}
\label{MdMpl}
\end{figure*}

\begin{figure*}
\centerline{\Huge \underline{Permissible disc masses and constituent disc particle radii}}
\centerline{}
\centerline{}
\centerline{
\includegraphics[width=8.5cm]{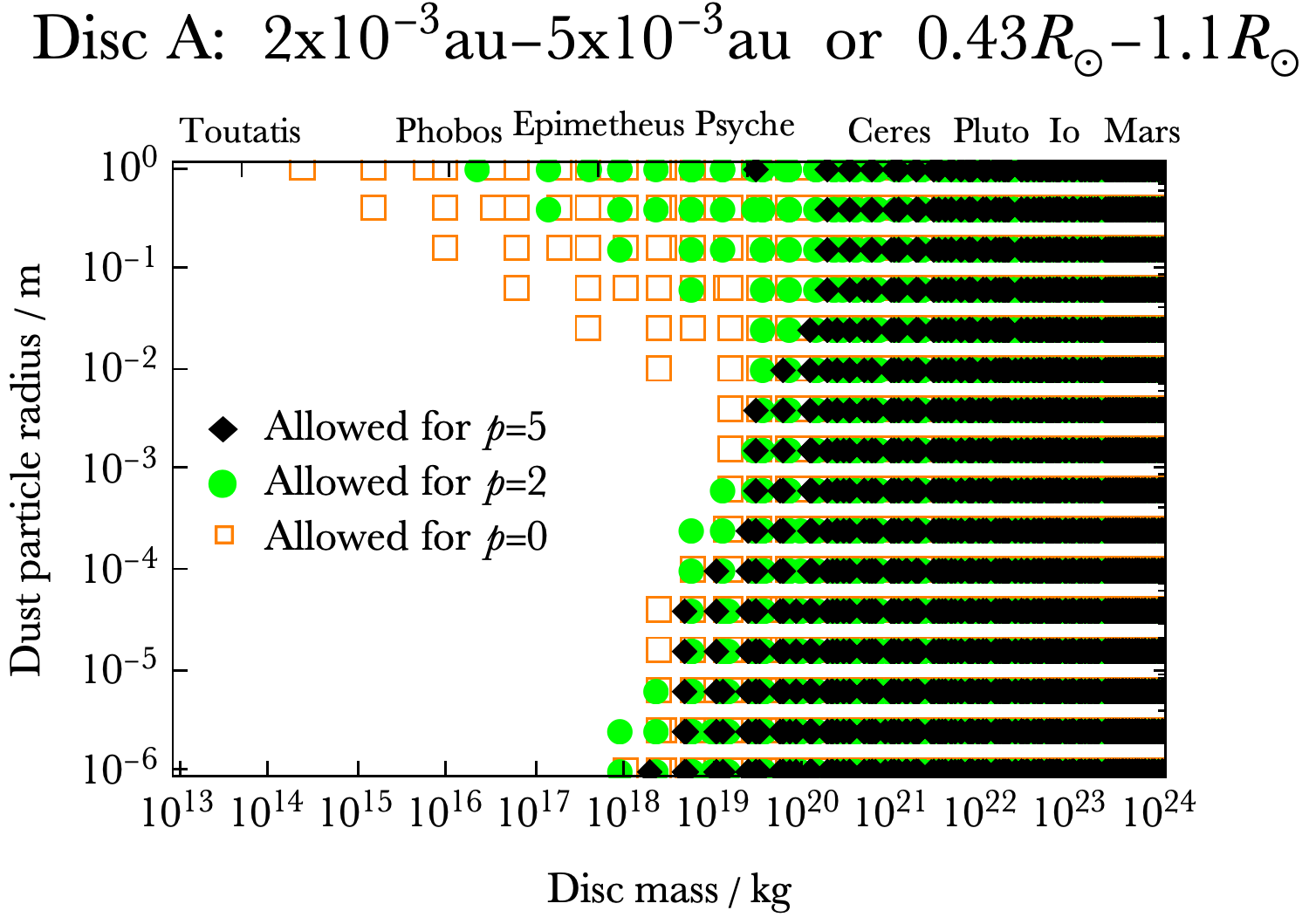}
\includegraphics[width=8.5cm]{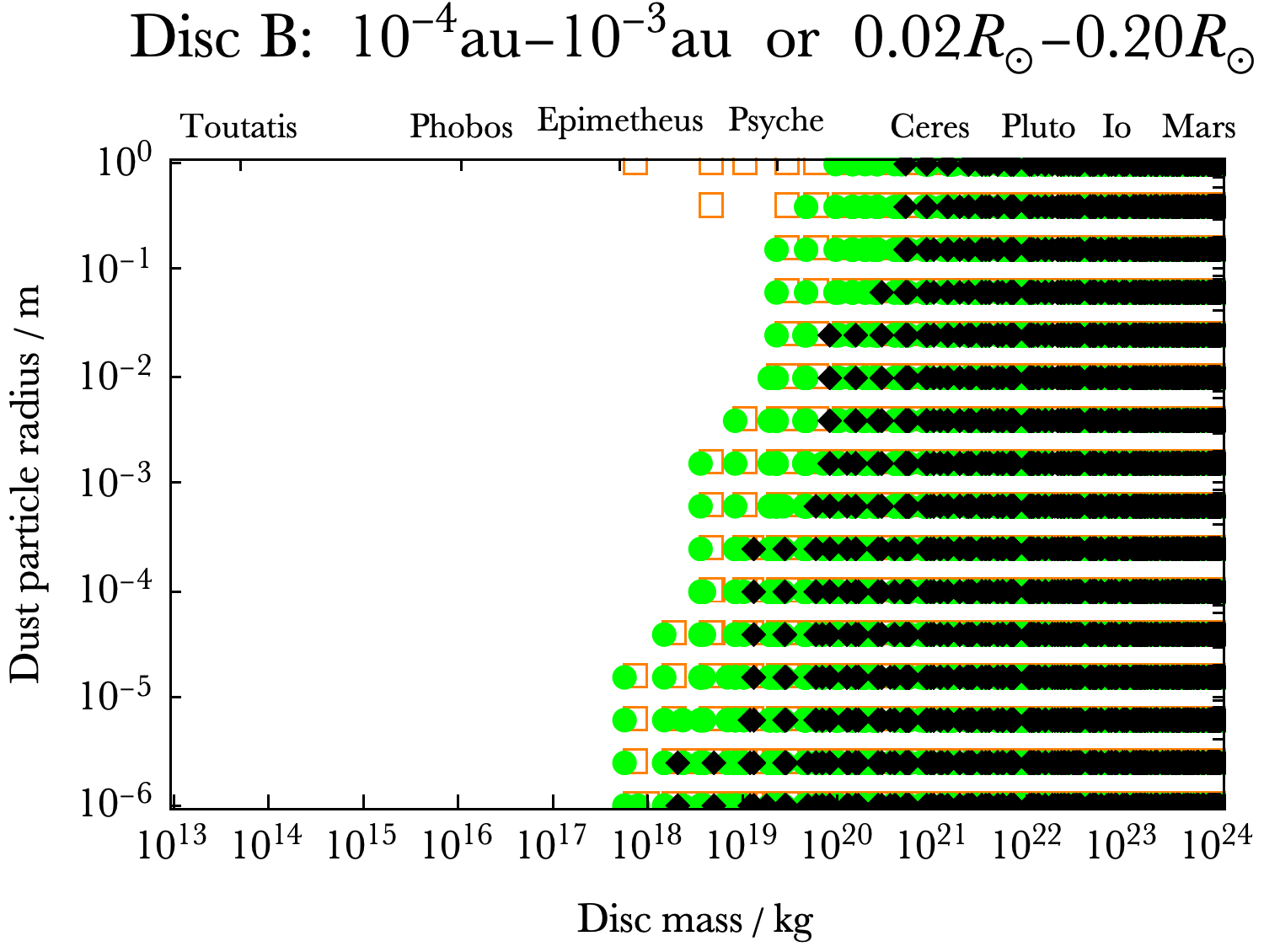}
}
\centerline{}
\centerline{
\includegraphics[width=8.5cm]{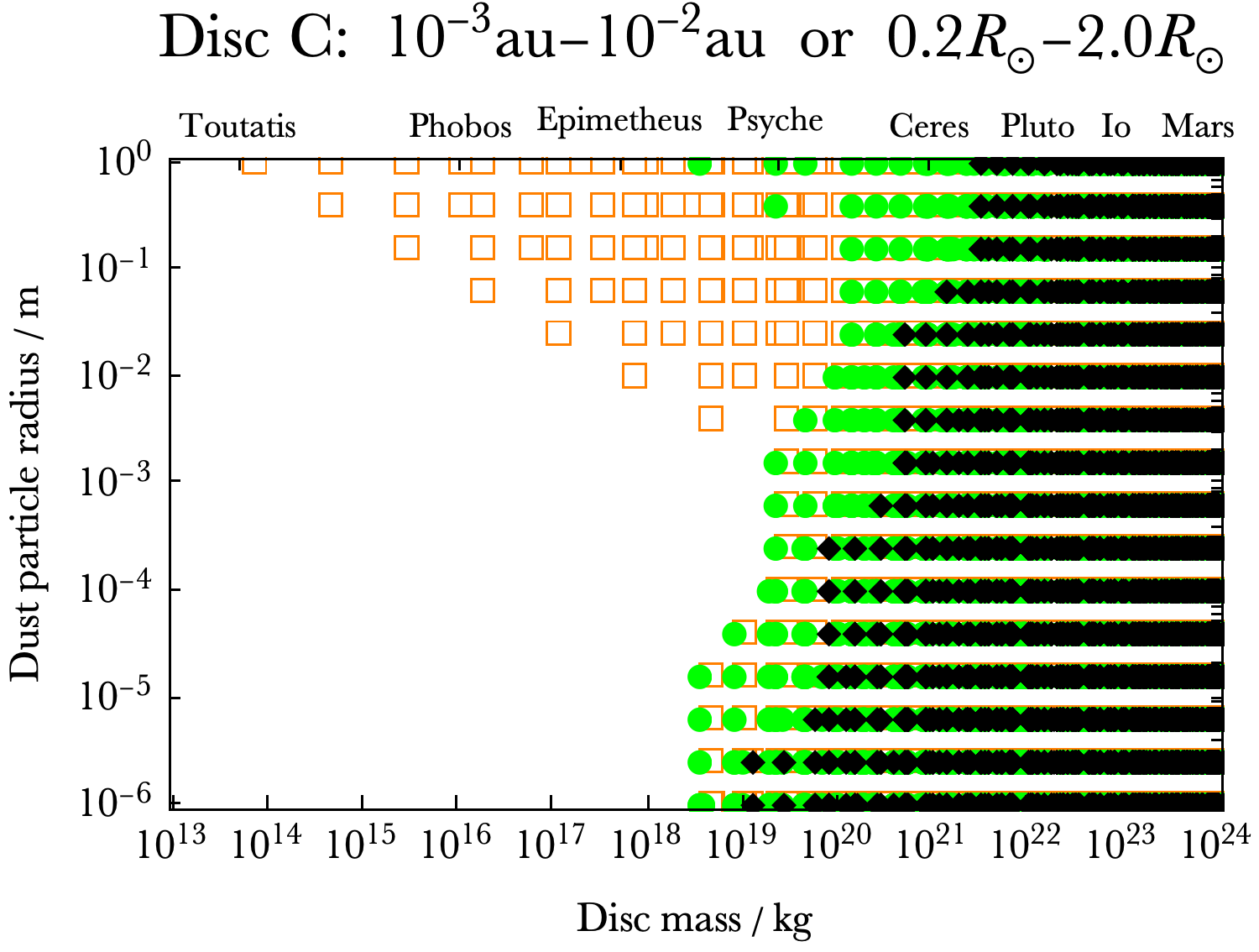}
\includegraphics[width=8.5cm]{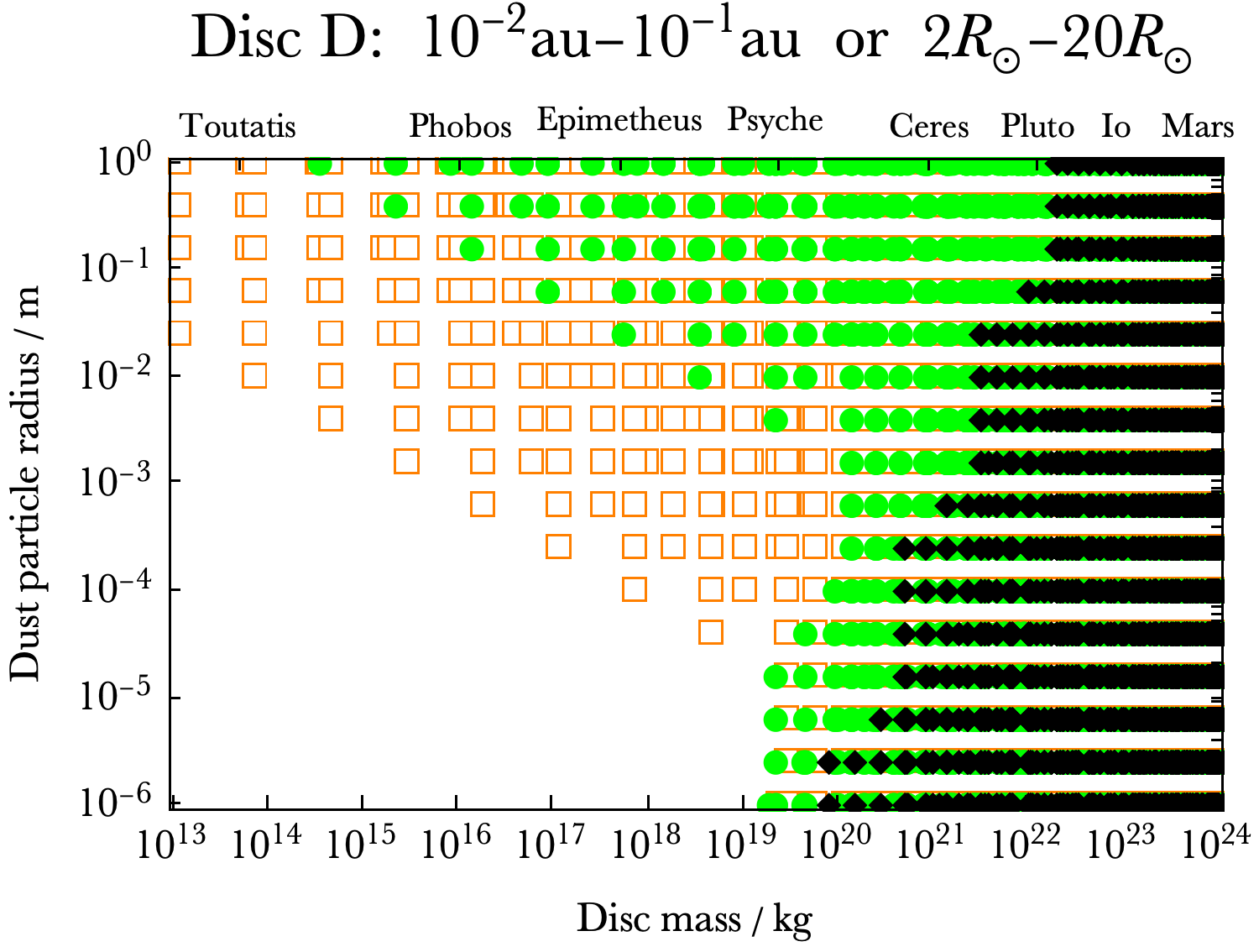}
}
\caption{
Combinations of disc masses and disc particle radii which allow for Type III migration to occur in dusty parts of white dwarf debris discs. Each symbol represents a permissible set of all {\revv 11} input variables; see Section 4.5. The absence of a symbol indicates a region of parameter space that is not allowed. The entire range of disc particle radii which were sampled are allowed for disc masses $\gtrsim 10^{19}$~kg.
}
\label{MdRpart}
\end{figure*}

\begin{figure*}
\centerline{\Huge \underline{Permissible disc masses and dynamical excitation level}}
\centerline{}
\centerline{}
\centerline{
\includegraphics[width=8.5cm]{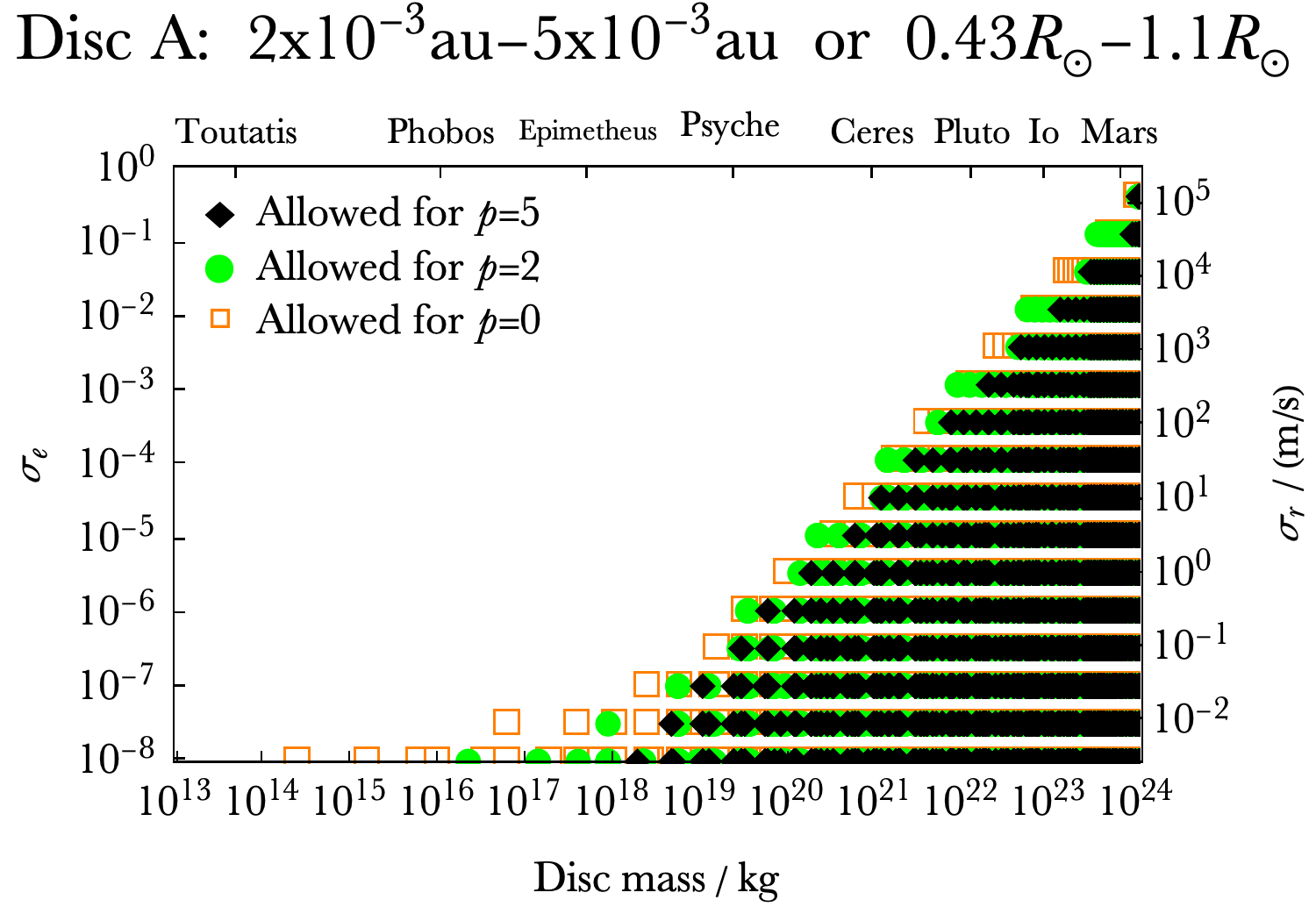}
\includegraphics[width=8.5cm]{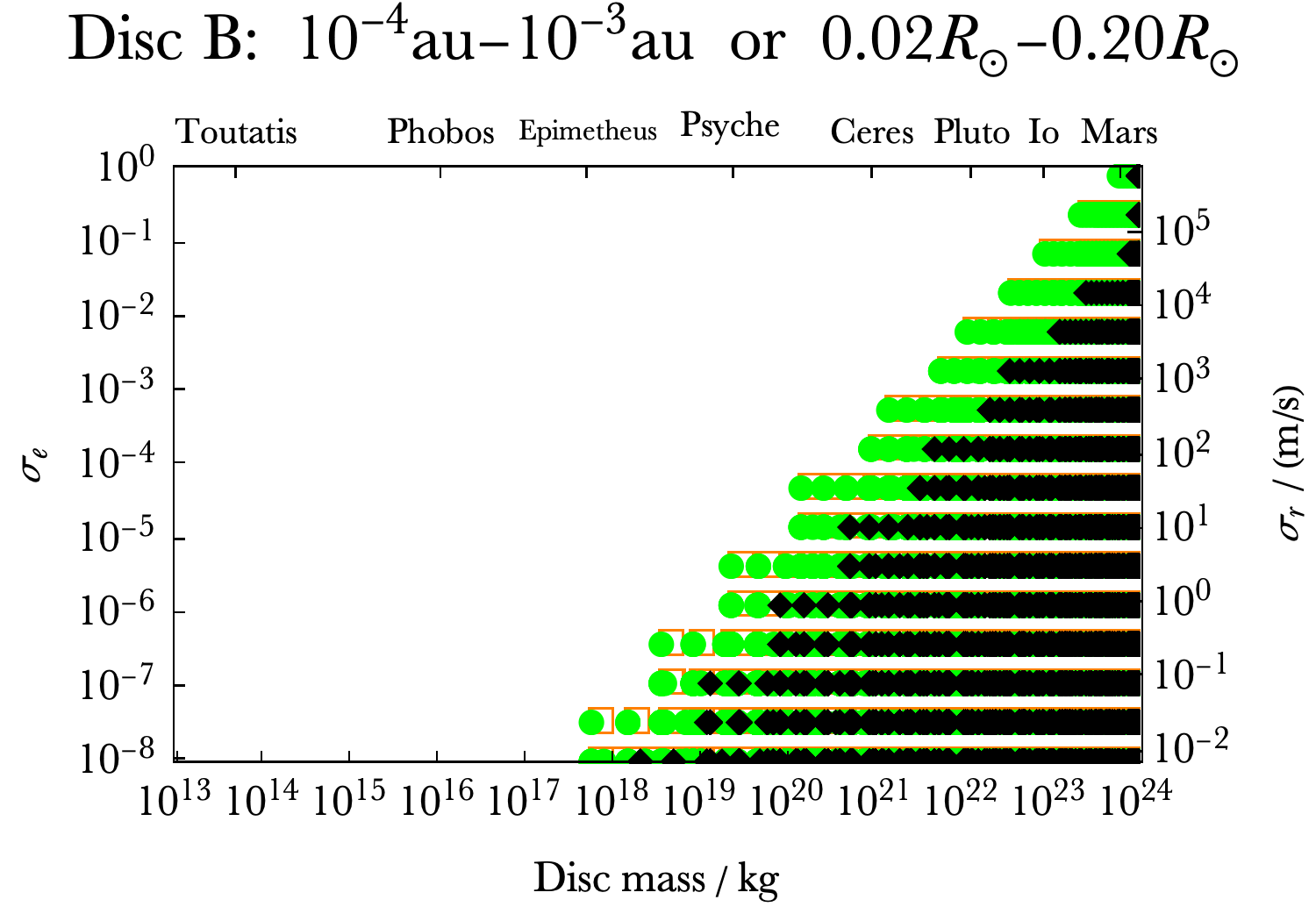}
}
\centerline{}
\centerline{
\includegraphics[width=8.5cm]{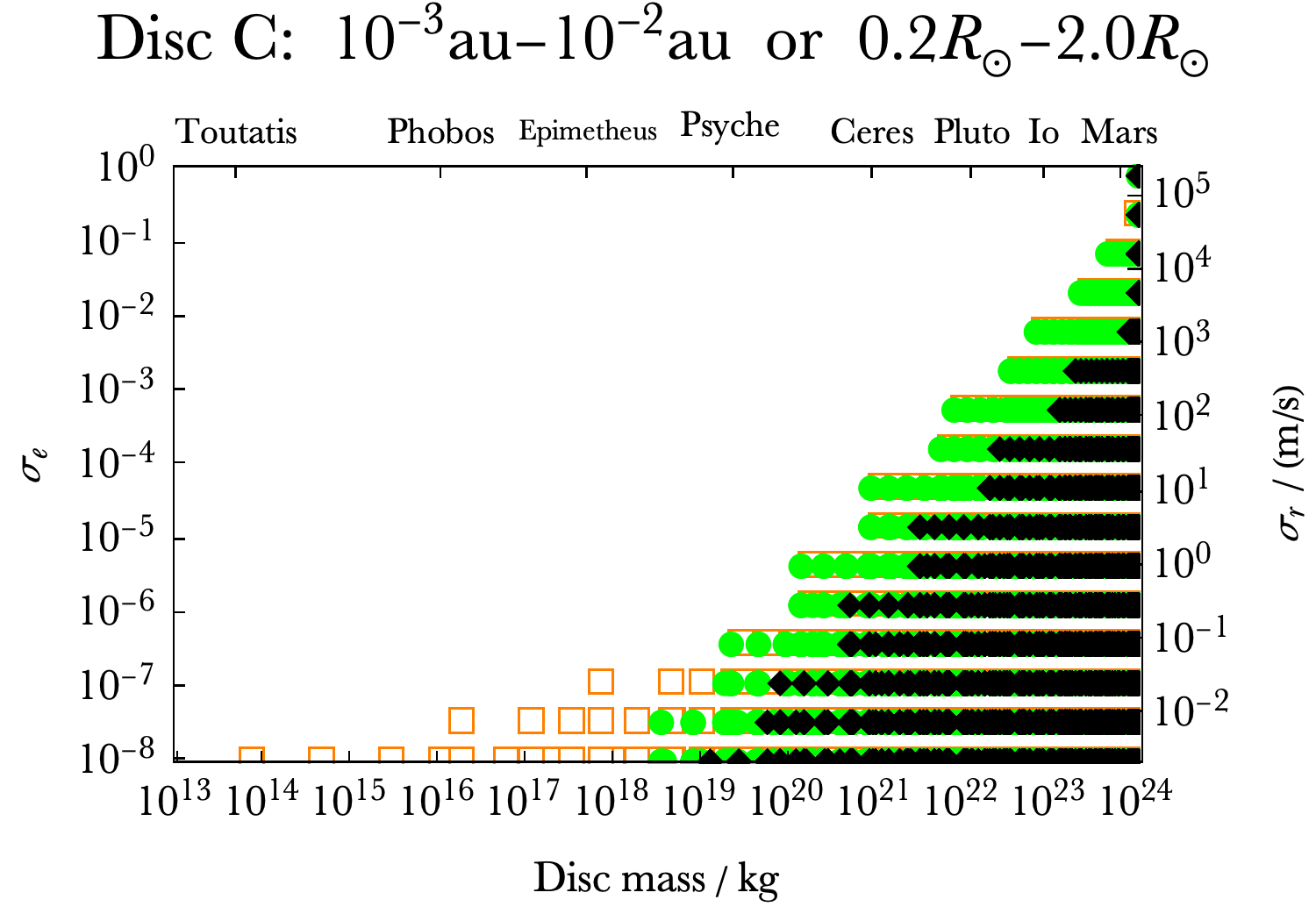}
\includegraphics[width=8.5cm]{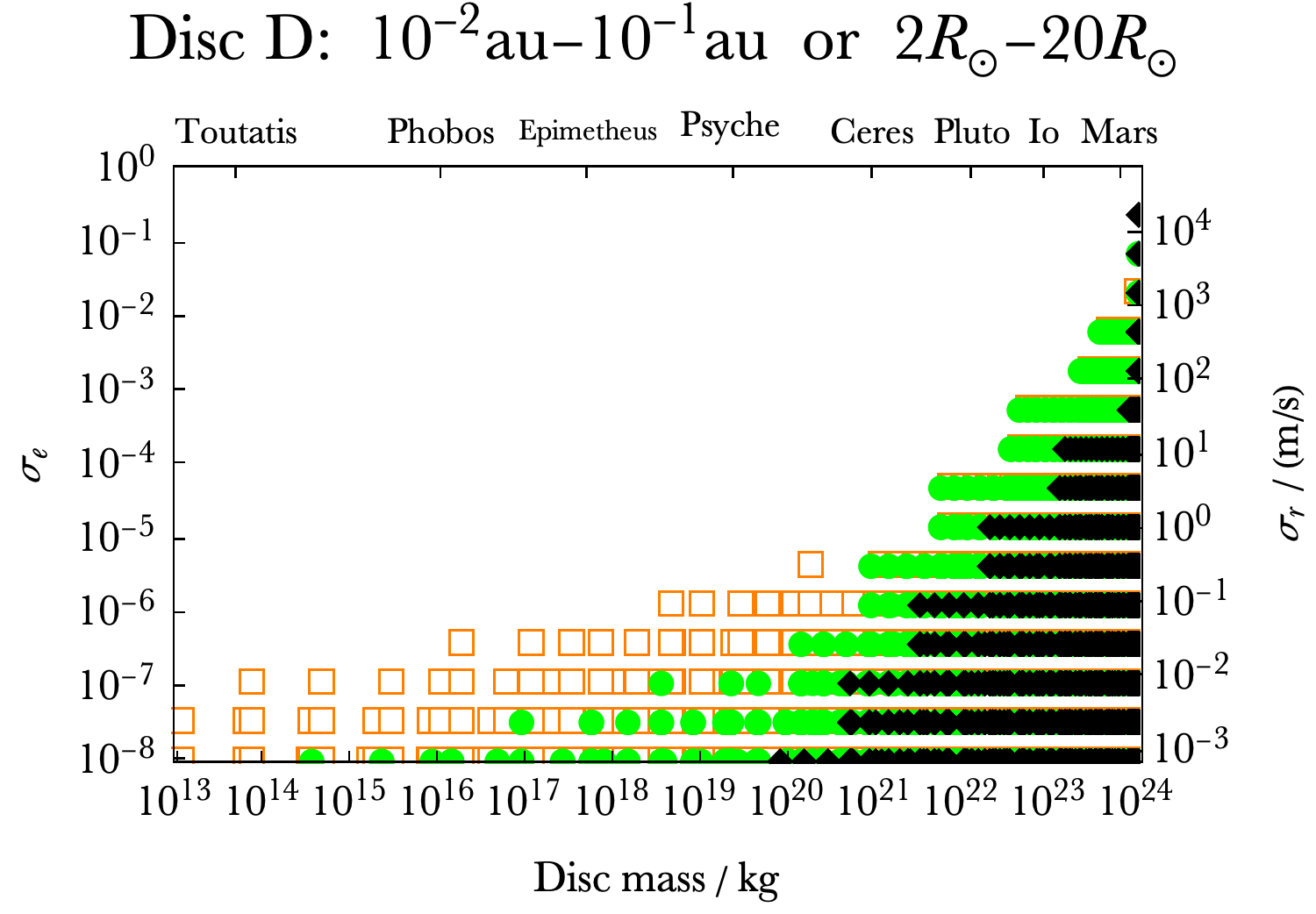}
}
\caption{
The dynamical excitation of a white dwarf debris disc that would allow for Type III planetesimal migration to occur in dusty regions. The left $y$-axis shows the root-mean-squared eccentricity of disc particles, and the right $y$-axis displays their radial velocity dispersion. Each symbol represents a permissible set of all {\revv 11} input variables; see Section 4.5. The absence of a symbol indicates a region of parameter space that is not allowed. The least-excited discs provide the largest area of parameter space which would allow for planetesimals to migrate. 
}
\label{Mdsigmae}
\end{figure*}

\begin{figure*}
\centerline{\Huge \underline{Permissible planetesimal masses and vertical scale heights}}
\centerline{}
\centerline{}
\centerline{
\includegraphics[width=8.5cm]{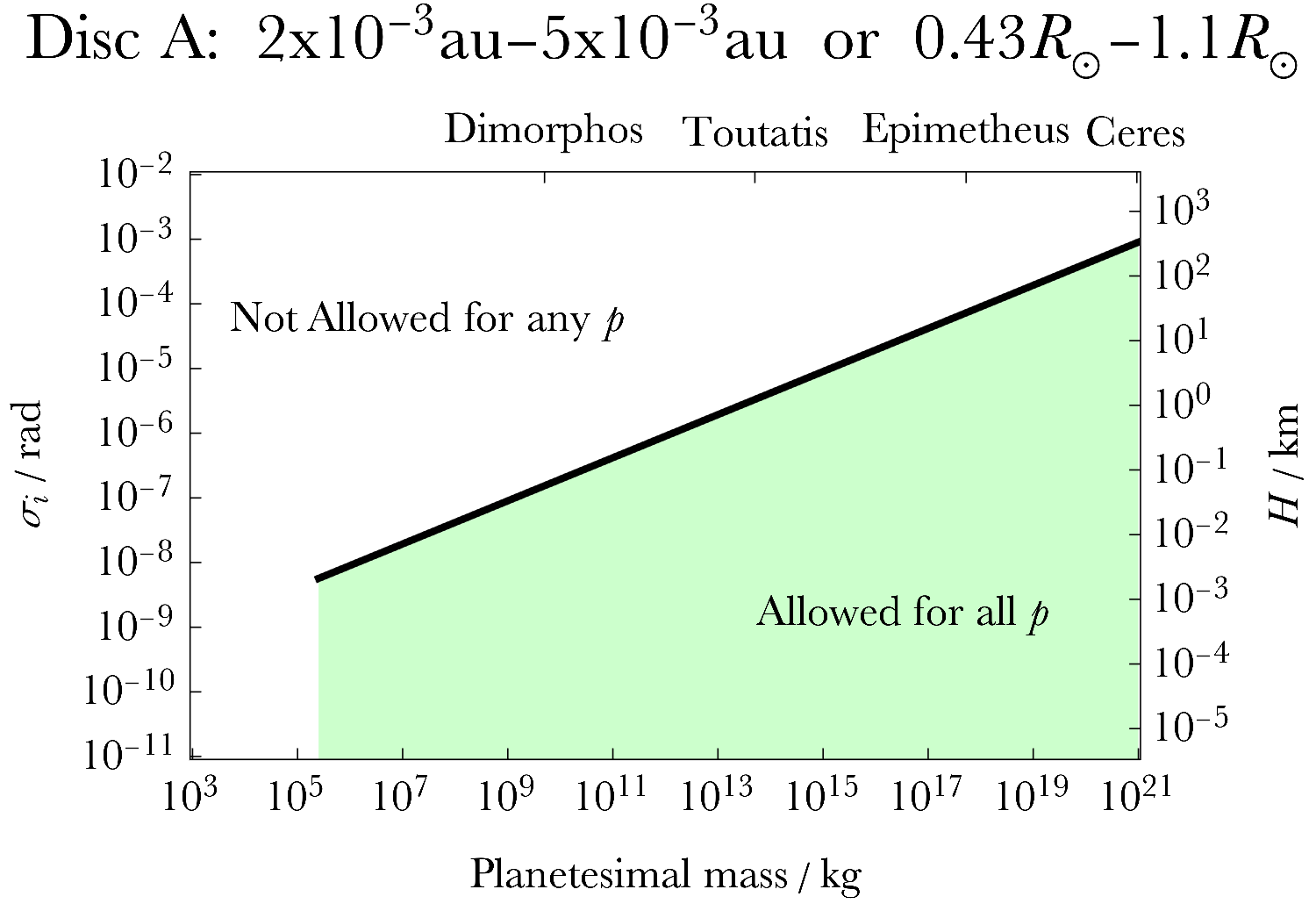}
\includegraphics[width=8.5cm]{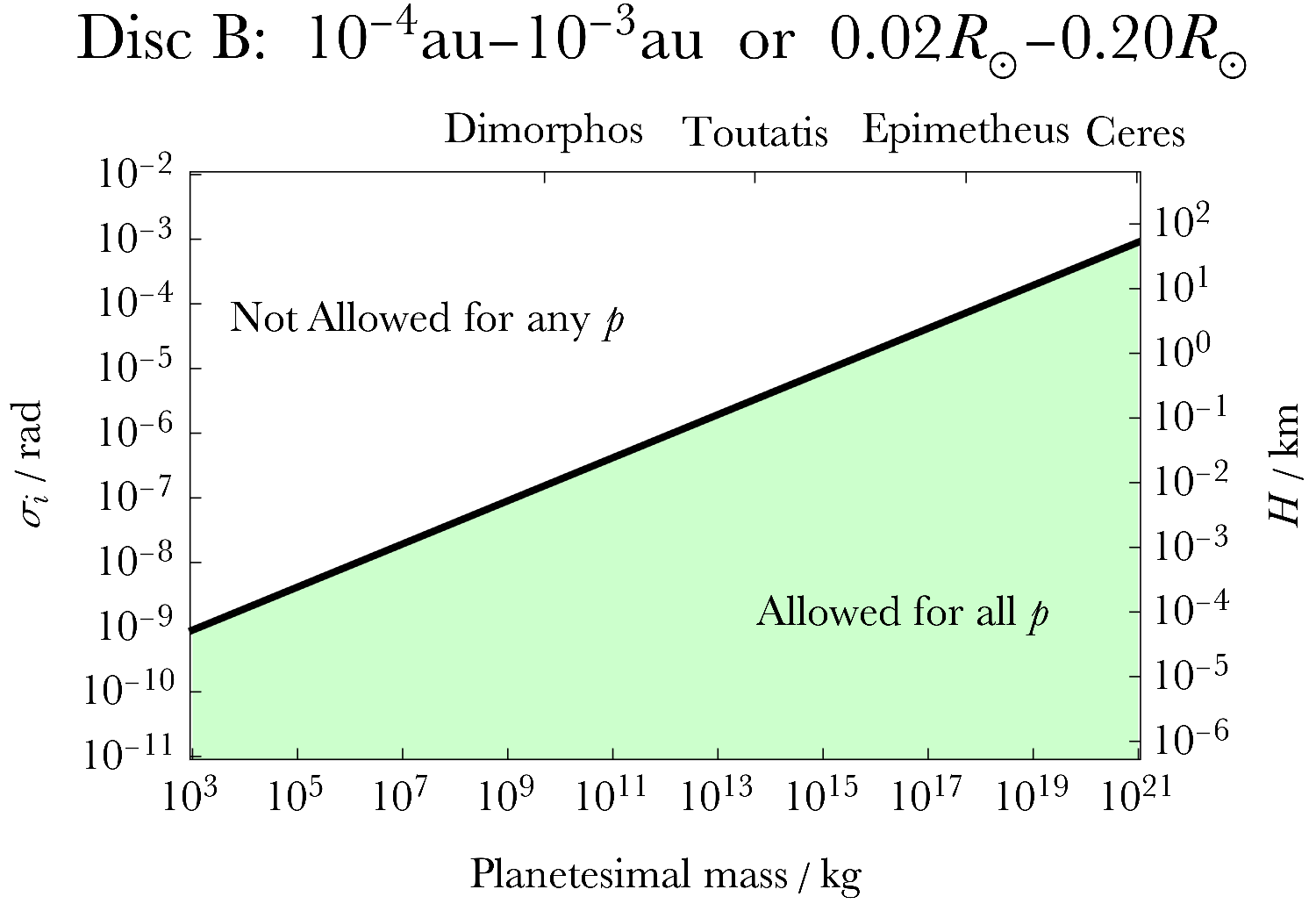}
}
\centerline{}
\centerline{
\includegraphics[width=8.5cm]{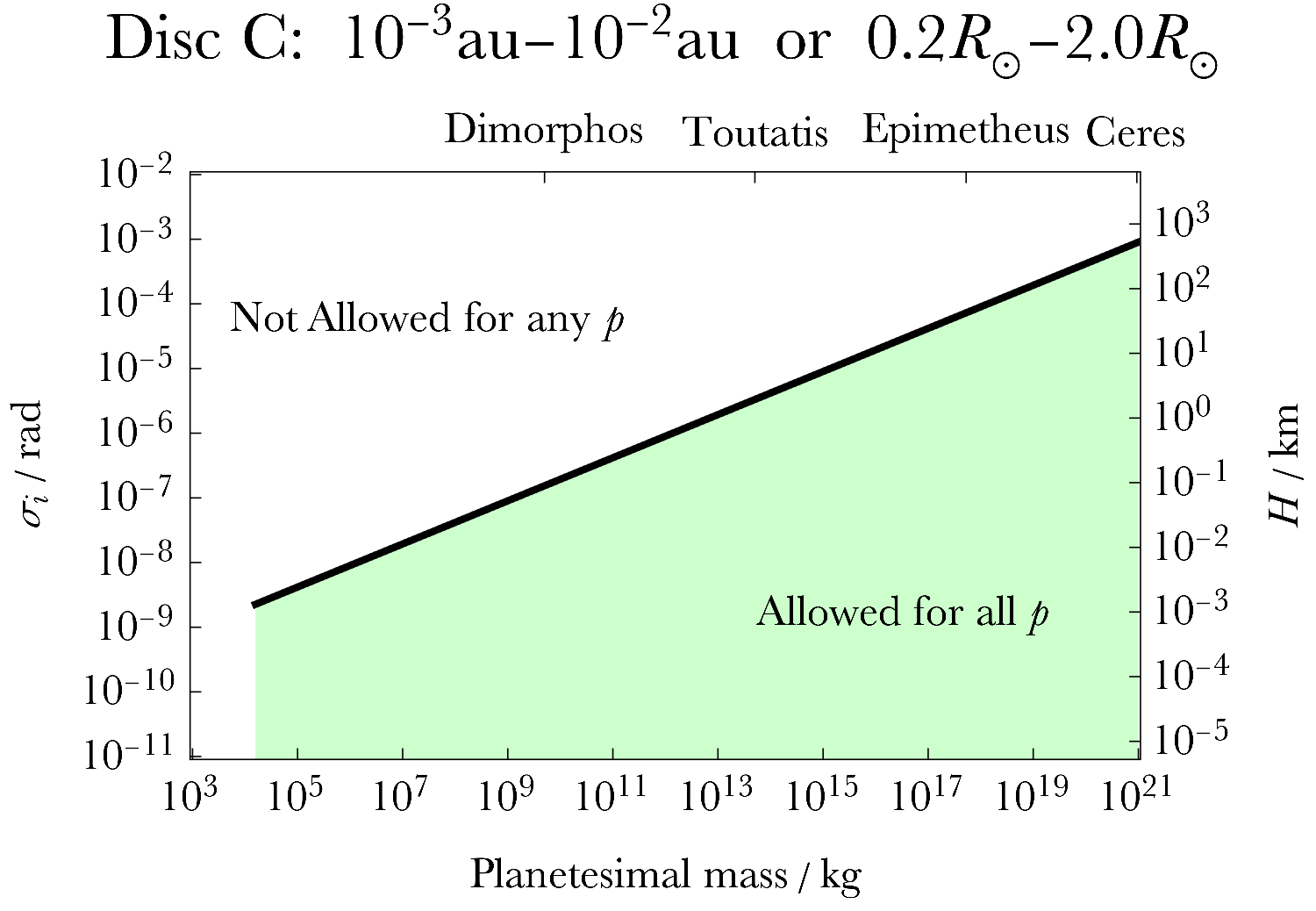}
\includegraphics[width=8.5cm]{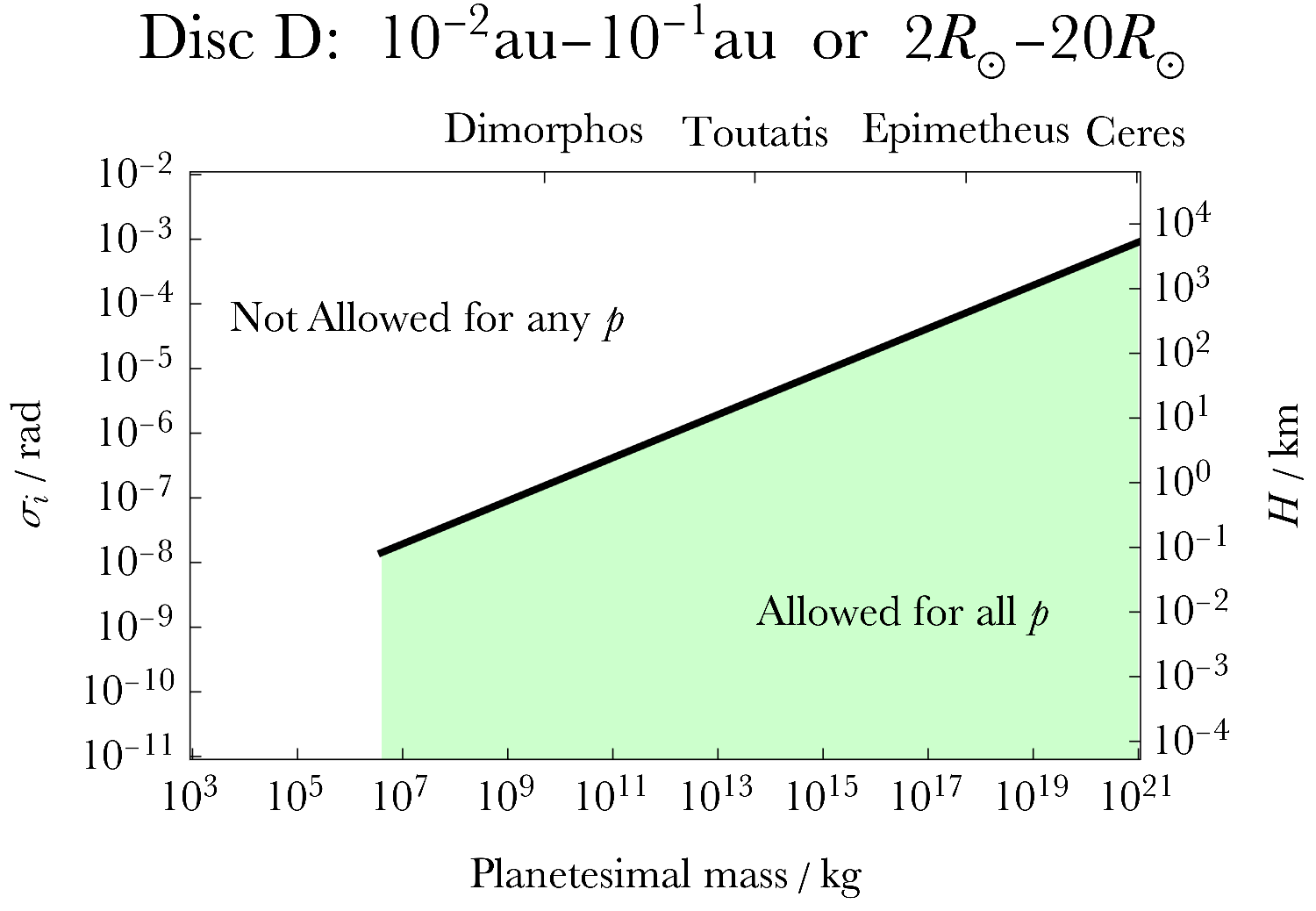}
}
\caption{
Combinations of planetesimal masses, root-mean-squared inclination of disc particles (left $y$-axis), and vertical scale heights (right $y$-axis) which allow for Type III migration to occur in dusty parts of white dwarf debris discs. The permissible disc scale heights are tied to the planetesimal's Hill radius through its mass.
}
\label{Mplsigmai}
\end{figure*}

\subsection{Conditions to activate Type III migration}

Having shown that Type III migration is potentially effectual, we now quantify in detail when, and if, this regime can be active in the first place for white dwarf debris discs. As mentioned in Section 4.3, three relations must be satisfied ($r_{\rm Hill} > H$, $r_{\rm Hill} < r_{\rm fast}(r)$, $r_{\rm Hill} > r_{\rm gap}(r)$) for some set of values of our {\revv 11 initial parameters ($r_{\rm in}$, $r_{\rm out}$, $M_{\star}$, $M_{\rm d}$, $M_{\rm pl}$, $p$, $\sigma_e$, $\sigma_i$, $R_{\rm part}$, $\rho_{\rm part}$, $a_{\rm pl}$)}.

To help decrease the parameter space to explore, we can identify five variables to fix ($r_{\rm in}$, $r_{\rm out}$, $M_{\star}$, $\rho_{\rm part}$, $a_{\rm pl}$) as follows:

\begin{itemize}

\item Our four predefined discs set $r_{\rm in}$ and $r_{\rm out}$.

\item We simply set $a_{\rm pl} = a_{\rm d}$, which is located in-between $r_{\rm in}$ and $r_{\rm out}$. 

\item We fix $M_{\star} = 0.65M_{\odot}$ and $\rho_{\rm part} = 2\, $g\,cm$^{-3}$ because these values will not vary significantly enough to substantially affect the final results.

\end{itemize}

Further, we do not need to vary $\sigma_i$, because it enters only into the condition $r_{\rm Hill} > H$ and is bound by the other variables through

\begin{equation}
\sigma_i < \frac{2\sqrt{2} a_{\rm pl}}{r_{\rm out}+r_{\rm in}}
           \left( \frac{M_{\rm pl}}{3M_{\star}} \right)^{1/3}
           .
\end{equation}

Hence, we focus on {\revv varying just $M_{\rm d}$, $M_{\rm pl}$, $p$, $\sigma_e$, and $R_{\rm part}$} in order to explore for what parameter sets Type III migration is possible. First, $\sigma_e$ is naturally bound by

\begin{equation}
0 < \sigma_{e} < \frac{r_{\rm out} - r_{\rm in}}{r_{\rm out} + r_{\rm in}}.
\end{equation}

\noindent{}The idealised case of $\sigma_{e} = 0$ yields an inviscid disc which always satisfies $r_{\rm Hill} > r_{\rm gap}(r)$. We set min$(\sigma_e) = 10^{-8}$ as a generous lower limit; this value corresponds to about $\sigma_r = $1~mm/s in the most extreme case of Disc D.  

Then, we explore the generous range {\rev $M_{\rm pl} = 10^3-10^{21}$~kg}, where the upper bound is approximately equivalent to a Ceres mass. Fixing $M_{\rm pl}$ in turn allows us to bound $M_{\rm d}$ from below {\revv by using the condition} $r_{\rm Hill} < r_{\rm fast}(r)$, {\revv generating the second parts of the following arguments:}

\[
M_{\rm d} > 
\]

\[
{\rm max}\left[\left[ \frac{r_{\rm out}^{2-p} - r_{\rm in}^{2-p}}{2\left(2-p\right)r_{\rm in}^{2-p}} \right] M_{\rm pl}, \ 
             1.04 \left( \frac{ r_{\rm out}^{2-p} - r_{\rm in}^{2-p} }{2-p} \right) \frac{r_{\rm out}^{p} M_{\star}^{\frac{1}{3}}M_{\rm pl}^{\frac{2}{3}}}{a_{\rm pl}^2}
                     \right], 
\]

\[                     
\ \ \ \ \ \ p < 2;
\]

\[
{\rm max}\left[\frac{1}{2} \ln{\left(\frac{r_{\rm out}}{r_{\rm in}} \right)} M_{\rm pl}, \ 
             1.04 \ln\left( \frac{ r_{\rm out} }{ r_{\rm in} } \right) \frac{r_{\rm out}^{p} M_{\star}^{\frac{1}{3}}M_{\rm pl}^{\frac{2}{3}}}{a_{\rm pl}^2}
                     \right], 
\]                     
  
\[                     
\ \ \ \ \ \ p = 2;
\]

\[
{\rm max}\left[\left[ \frac{r_{\rm out}^{2-p} - r_{\rm in}^{2-p}}{2\left(2-p\right)r_{\rm out}^{2-p}} \right] M_{\rm pl}, \ 
             1.04 \left( \frac{ r_{\rm out}^{2-p} - r_{\rm in}^{2-p} }{2-p} \right) \frac{r_{\rm out}^{p} M_{\star}^{\frac{1}{3}}M_{\rm pl}^{\frac{2}{3}}}{a_{\rm pl}^2}
                     \right], 
\]

\begin{equation}                     
\ \ \ \ \ \ p > 2,
\end{equation}

\noindent{}where we have {\revv also imposed an additional restriction in the first part of each argument about the mass ratio of the planetesimal relative to that of the disc. If this mass ratio is too small, migration will not occur \citep{broken2011}. To derive these arguments, we adopted the local surface density conditions which set this minimum mass to be $M_{\rm pl} \lesssim 4\pi\Sigma_{\rm d}a_{\rm pl}^2$ \citep{frietal2022}}.

Finally, for $R_{\rm part}$, as previously mentioned, we do have evidence of the presence of micron-sized grains and the breakup of asteroids, but little information about sizes in-between these extremes. Hence, we sample a generous range of $R_{\rm part} = 10^{-6} - 10^{-0}$~m. Fixing $R_{\rm part}$ finally allows us to perform just a single test on the {\revv remaining condition ($r_{\rm Hill} > r_{\rm gap}(r_{\rm in})$)} in order to determine if Type III migration is possible.

We present the results of these tests in Figs. \ref{MdMpl}-\ref{Mplsigmai}, where each figure consists of four plots, one per disc (A through D). On Figs. \ref{MdMpl}-\ref{Mdsigmae}, the presence of {\revv symbols indicates} permissible region of parameter space {\revv for given values of $p$}, and the absence of symbols indicates a disallowed region. For this reason, some plots contain a significant amount of whitespace. {\rev Alternatively, in Fig. \ref{Mplsigmai} a range of permissible values are shaded in.} Within the ranges specified above, each variable was sampled {\revv 15} times in a logarithmic interval. Also indicated on at least one axis of each plot are masses of particular solar system objects, for context.

The shapes of the structures defined by the points are determined by the combined solution of the three inequalities which set the conditions for Type III migration. {\rev However, note that with {\revv five} input variables (the other six being fixed), the solutions are degenerate. Hence, drawing analytical curves on the plots which bound the shapes is not possible without making additional assumptions.} 

These shapes provide valuable constraints on pairs of parameters, as presented here. For example, Fig. \ref{MdMpl} reveals that only discs more massive than those created by Epimetheus-mass planetesimals can effectively host dusty Type-III migration within $10^{-2}$ au. Further, within these discs, the range of potential masses of planetesimals that migrate increases with increasing disc mass, and the most likely migrator would have a mass similar to Ryugu. {\rev Further, planetesimal masses less than $10^5$~kg can migrate only within the most extended and distant discs.}

Figure \ref{MdRpart} is arguably not as constraining with respect to the disc particle radius. Every one of the four discs can admit a migrator for any of the disc particle radius values sampled. Perhaps notable is that Disc B, representing the most compact disc, has the least capability to host a migrator for the largest disc particle radii. The more useful constraint is again on the disc mass.

Figure \ref{Mdsigmae} provides a sense of how dynamically excited the disc may be in the azimuthal and radial directions and still host a migrator. Note that the left $y$-axis provides values for $\sigma_e$ and the right $y$-axis provides values for $\sigma_r$. The plots show that the most symmetrical discs provide the widest range of disc masses that would allow for planetesimals to migrate, and roughly support our previous claim that disc masses smaller than an Epimetheus-mass are unlikely to support migration.


Finally, Fig. \ref{Mplsigmai} displays permissible values for both $\sigma_i$ and $H$. These values are pinned to the planetesimal mass through the definition of Hill radius and the condition $r_{\rm Hill} > H$, {\rev which is why we can shade the permissible regions, as opposed to individual points, on these plots}. These plots provide strict coupled constraints on the vertical extent of a disc and its capability to host a migrating planetesimal, and {\rev demonstrate that migration within very thin discs similar to Saturn's rings \citep{jura2003,rafikov2011a,rafikov2011b} is possible.} If scale heights can be better observationally constrained, then those constraints would enable us to exclude a population of planetesimals from being able to migrate.

\subsection{Migration timescales}

Having now identified where in parameter space Type III migration is possible, we may explicitly solve equation (\ref{dadtMigIII}) for $a_{\rm pl}(t)$. Doing so requires us to assume that {\revv 10 of our 11} input variables are independent of time, with the exception being $a_{\rm pl}$. The result is



\begin{equation}
a_{\rm III}(t) = \left[ a_{\rm III,i}^{-1/2} \mp 
                     1.3 \frac{\mathcal{G}^{1/2} \Sigma_{\rm d}}
                          {M_{\star}^{1/2}} t \right]^{-2}.
\label{aIII}
\end{equation}

Therefore, we can invert equation (\ref{aIII}) to solve explicitly for the timescale over which a planetesimal would migrate across the entire disc, $\tau_{\rm traversal}^{\left({\rm Whole \, disc}\right)}$. Assuming that the nature of migration is inward, and that throughout the migration $r_{\rm Hill}(t)$, $r_{\rm gap}(t)$ and $r_{\rm fast}(t)$ do not change sufficiently enough to alter the requirements for Type III migration to occur, we obtain

\begin{figure}
\centerline{
\includegraphics[width=8.5cm]{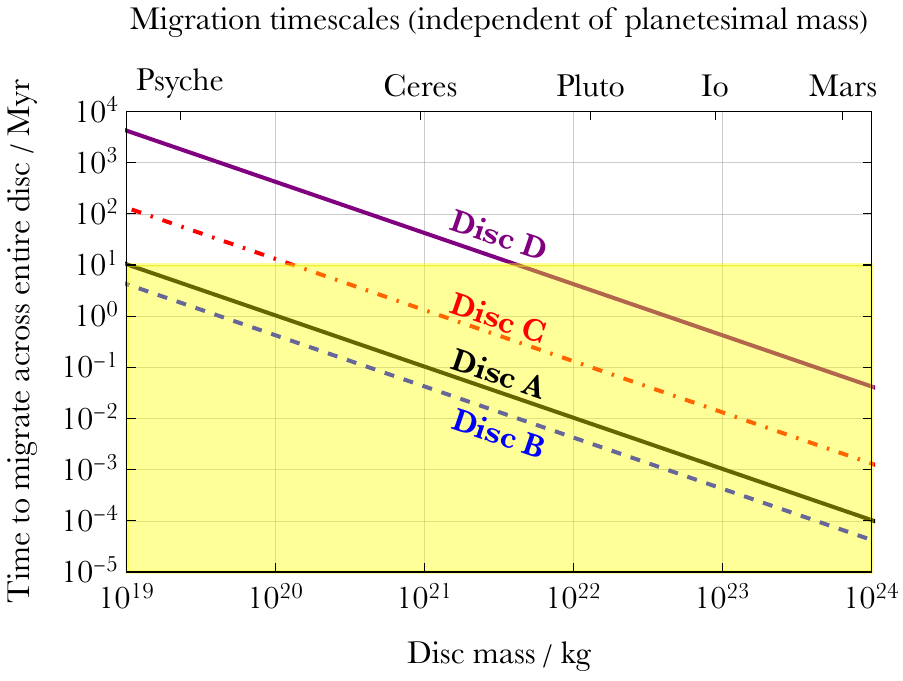}
}
\caption{
Timescale for a planetesimal of any mass (less than that of the disc) to migrate in a Type III fashion across the entire disc from its outer edge to its inner edge, as a function of disc mass. The conditions to allow for migration are assumed to be satisfied throughout the migration. {\revv The yellow shaded region indicates the range of realistic disc lifetimes.}
}
\label{DiscTraversal}
\end{figure}

\begin{figure}
\centerline{
\includegraphics[width=8.5cm]{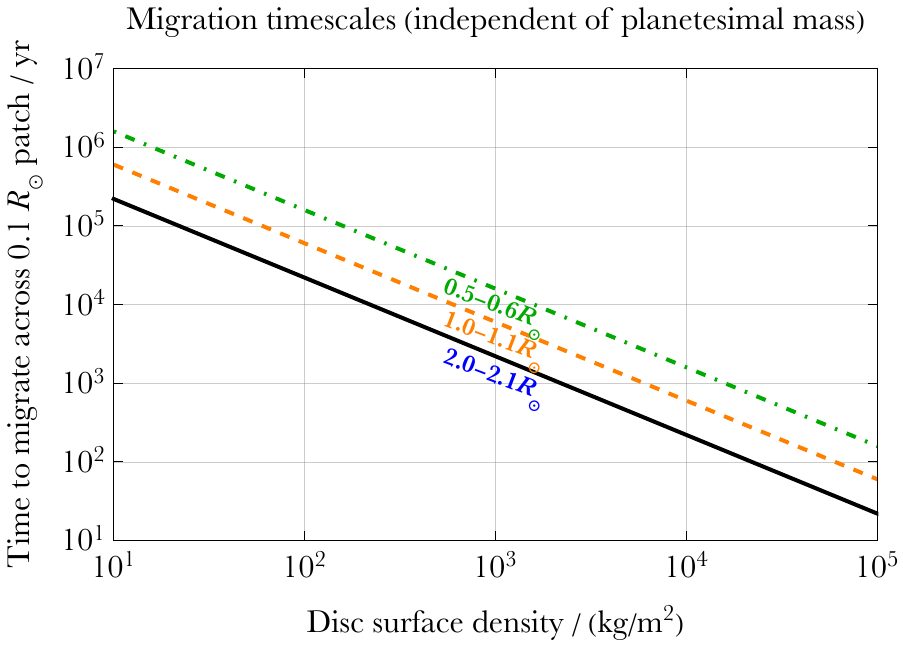}
}
\caption{
Timescale for a planetesimal of any mass (less than that of the disc) to migrate in a Type III fashion across a $0.1R_{\odot}$ patch of a disc as a function of surface density, for three patches at different distances from the white dwarf.
}
\label{PatchTraversal}
\end{figure}

\begin{equation}
\tau_{\rm traversal}^{\left({\rm Whole \, disc}\right)} \approx 
\frac{2.4 M_{\star}^{1/2}}{\mathcal{G}^{1/2} M_{\rm d}}
\left(r_{\rm out}^2 - r_{\rm in}^2\right)
\left(r_{\rm in}^{-1/2} - r_{\rm out}^{-1/2} \right).
\label{traverse}
\end{equation}

{\rev How robust is the assumption that $r_{\rm Hill}(t)$, $r_{\rm gap}(t)$ and $r_{\rm fast}(t)$ will satsify the conditions for the Type III regime throughout the migration, especially since each of those three quantities has a different functional dependence on $a_{\rm pl}$? The answer is highly degenerate because it depends on the choices for all {\revv 11} input variables.}

{\rev 
However, we can obtain a representative idea by comparing the four plots on each of the figures {\revv from Figs. \ref{MdMpl}-\ref{Mdsigmae}}. In each figure, all four plots have the same axes ranges, but represent different values of $r_{\rm in}$ and $r_{\rm out}$. Hence, comparing which points are common to all four plots suggests parameter space cross-sections that can admit migration across different radial scales.  
}

{\rev
Individual known systems would require dedicated treatments, but here we can provide some order-of-magnitude estimates.} Applying equation (\ref{traverse}) to our four pre-defined discs yields

\begin{equation}
\tau_{\rm traversal}^{\left({\rm Disc \, A}\right)} \approx
0.11 \, {\rm Myr} 
\left( \frac{M_{\rm d}}{M_{\rm Ceres}} \right)^{-1}
\left( \frac{M_{\star}}{0.65 M_{\odot}} \right)^{1/2}
,
\label{travA}
\end{equation}

\begin{equation}
\tau_{\rm traversal}^{\left({\rm Disc \, B}\right)} \approx
0.045 \, {\rm Myr} 
\left( \frac{M_{\rm d}}{M_{\rm Ceres}} \right)^{-1}
\left( \frac{M_{\star}}{0.65 M_{\odot}} \right)^{1/2}
,
\end{equation}

\begin{equation}
\tau_{\rm traversal}^{\left({\rm Disc \, C}\right)} \approx
1.4 \, {\rm Myr} 
\left( \frac{M_{\rm d}}{M_{\rm Ceres}} \right)^{-1}
\left( \frac{M_{\star}}{0.65 M_{\odot}} \right)^{1/2},
\end{equation}

\noindent{}and

\begin{equation}
\tau_{\rm traversal}^{\left({\rm Disc \, D}\right)} \approx
45 \, {\rm Myr} 
\left( \frac{M_{\rm d}}{M_{\rm Ceres}} \right)^{-1}
\left( \frac{M_{\star}}{0.65 M_{\odot}} \right)^{1/2}
.
\label{travD}
\end{equation}

The timescales in equations (\ref{travA})-(\ref{travD}) are all, importantly, independent of planetesimal mass. They also showcase an inversely proportional dependence on $M_{\rm d}$, and a dependence on $M_{\star}$ which is weak enough to ensure any variation in that variable produces a negligible shift in that timescale. Hence, we plot these four timescales as a function of $M_{\rm d}$ in Fig. \ref{DiscTraversal}. These timescales range from $10^2$ to $10^{10}$ yr depending on a range of realistic disc masses. {\revv Shaded on the plot is the range of realistic disc lifetimes from both observational and theoretical constraints \citep{giretal2012,verhen2020}; see Section 5.1.}

However, what if the disc was not uniform enough to be described by the same surface density at each location, as suggested by Doppler tomograms such as those in \cite{manetal2016,manetal2021}? In this case we can focus on an individual patch of the disc with a given surface density. We hence re-write equation (\ref{traverse}) as

\begin{equation}
\tau_{\rm traversal}^{\left({\rm Patch}\right)} \approx 
\frac{0.77 M_{\star}^{1/2}}{\mathcal{G}^{1/2} \Sigma_{\rm d}}
\left(a_{\rm patch,in}^{-1/2} - a_{\rm patch,out}^{-1/2} \right)
\end{equation}

\noindent{}and plot the result for three different patches which are $0.1R_{\odot}$ wide in radial extent in Fig. \ref{PatchTraversal}. This figure allows us to compare timescales for to migrate across equal distances, unlike in Fig. \ref{DiscTraversal}.

\section{Discussion}

Because Section 4 has demonstrated that disc-planetesimal-induced migration in white dwarf debris discs is possible, let us now explore its feasibility in the context of additional effects and observations in these systems.

\subsection{Disc lifetimes}

A fundamental constraint on the feasibility of migration is the disc lifetime. However, observational constraints are limited to discs most akin to Disc A and Disc C, and are obtained by dividing the metal masses contained within the envelopes of one type of white dwarf (DBZs) by the accretion rates measured in a different type of white dwarfs (DAZs) \citep{giretal2012}. The level of realism of this computation is unclear, but has yielded lifetimes of $0.03-5$ Myr. This computation also does not distinguish between gas-rich and dust-rich discs.

Theoretical computations of disc lifetime have been performed for particulate discs with $R_{\rm part} \ge 10^{-2}$~m \citep{verhen2020}. They predict that for $M_{\rm disc} = 10^{12}-10^{24}$~kg and $R_{\rm part} = 10^{-2}-10^{0}$~m, the extents of the discs studied here (Discs A, B, C and D) would not be able to maintain a steady state over 1 Myr, and in many cases, not even over 1 kyr. The most massive discs would not be able to maintain their architecture for even 1 yr, although in these cases collisional evolution would break the disc down into a different disc with smaller $R_{\rm part}$.

Further obstacles to maintain a long-lived steady state -- a disc environment which is most amenable to disc migration -- arises from both theoretical and observational considerations. \cite{kenbro2017a} showed how discs can oscillate between high and low states depending on how and when they are externally replenished. Further, {\revv infrared} flux changes within known white dwarf discs \citep{xujur2014,faretal2018b,xuetal2018,rogetal2020} appear to be ubiquitous \citep{swaetal2019}, and sometimes significantly large enough to indicate major collisional events {\rev \citep{wanetal2019,swaetal2021}}.

\subsection{Roche radii}

In order for a planetesimal to migrate inwards across a particular patch of a disc, the planetesimal must survive the migration and not enter its Roche radius. We compute representative Roche radii for several different planetesimals by using the same prescription as in \cite{veretal2022} and derived from \cite{holmic2008} and \cite{zhaetal2021}:

\begin{equation}
r_{\rm Roche} = \left[ 
            \left(\frac{8\pi}{15} + \frac{2\sqrt{3} S_{\rm pl}}{\mathcal{G} R_{\rm pl}^2 \rho_{\rm pl}^2} \right)
            \left( \frac{\rho_{\rm pl}}{M_{\star}} \right) 
                \right]^{-1/3},
\label{rroche}
\end{equation}

\noindent{}where $R_{\rm pl}$ and $\rho_{\rm pl}$ are the radius and density of the (assumed-to-be spherical) planetesimal and $S_{\rm pl}$ is the planetesimal's material cohesive strength.

We overplot the Roche radii for different planetesimals in Fig. \ref{DiscCartoon}. These different radii span nearly two orders of magnitude, from $10^{-4}$~au to $10^{-2}$~au, demonstrating different breakup limits for the planetesimal. Note in particular how the value of $S_{\rm pl}$ -- which does not enter in the Type III migration prescriptions or conditions -- can significantly affect the inner boundary of the possible migration region. 

Additionally, throughout this manuscript we have assumed that the planetesimal is homogeneous. If the planetesimal is composed of multiple layers -- and is large enough for example to contain a separate mantle and core -- then each layer will have its own Roche radius, such that some but not all of the planetesimal will survive when it passes one of its layer's Roche radii \citep{veretal2017,broetal2023a,broetal2023b}.

\subsection{Stellar tides}

Even if the planetesimal remains outside of the Roche radius, it may be affected by star-planet tides from the white dwarf. These tides can create their own migration mechanism (without the presence of a disc) that depends on the physical properties of the planetesimal, in addition to its total mass {\revv  \citep{veretal2019b,verful2019,ocolai2020,becetal2023}}. In particular, Fig. 3 of \cite{veretal2019b} suggests that (i) Ceres-mass planetesimals which have low internal dynamic viscosities of $\lesssim 10^{16}$ Pa$\cdot$s may migrate due to star-planet tides at comparable or faster rates than those induced by Type III migration, and (ii) lower-mass planetesimals would be more resistant to tidal migration, such that in these cases Type III migration would dominate.

{\rev Further, throughout this work we have assumed that all bodies, including disc particles and planetesimals, are spherical. We have also neglected their spin. Tides might change the shape and spin of planetesimals; if they rotate rapidly enough, they could mimic larger particles.}

\subsection{Sublimation radii}

Not only must inwardly migrating planetesimals survive break-up due to gravitational stretching, but they also must survive break-up by sublimating into gas. The prospects for sublimation depend strongly on the luminosity of the star, which in turn depends on the cooling age of the white dwarf\footnote{The cooling age of a white dwarf is defined as the length of time that the star has been a white dwarf.}.

We compute representative sublimation radii for arbitrary white dwarf cooling ages by using the same prescription as in \cite{veretal2022}, which was derived from \cite{mestel1952} and \cite{rafikov2011b}:

\begin{equation}
r_{\rm sub} = \frac{1}{4}
              \sqrt{\frac{3.53 L_{\odot}}{\pi \sigma T_{\rm sub}^4}
                    \left( \frac{M_{\star}}{0.65 M_{\odot}} \right)  
                    \left(0.1 + \frac{t_{\rm cool}}{\rm Myr} \right)^{-1.18} }
\label{rsub}
\end{equation}

\noindent{}where $L_{\odot}$ is the Sun's luminosity, $t_{\rm cool}$ is the white dwarf's cooling age, and $T_{\rm sub}$ is the sublimation temperature of a given material; for graphite, $T_{\rm sub} \approx 2600$~K and for iron, $T_{\rm sub} \approx 1600$~K \citep{rafgar2012}.

We solve equation (\ref{rsub}) for different white dwarf cooling ages and overplot the results on Fig. \ref{DiscCartoon}. Just like for the Roche radii, the sublimation radii extend from $10^{-4}$~au to $10^{-2}$~au, demonstrating that for the oldest white dwarfs, migrating planetesimals would be resistant to sublimation until the planetesimals reach distances of $\sim 10^{-4}$~au.

{\rev This treatment of sublimation is admittedly simplistic because large planetesimals sublimate over a non-zero timescale \citep{steetal2021}. Further, planetesimals are not necessarily homogeneous compositionally; some parts of the planetesimals will sublimate quickly, and others not at all before being broken up through tidal forces.}

\subsection{Radiative drag}

Radiation which is not luminous enough to sublimate a planetesimal may still drag it. Further, radiation plays a key role in the evolution of the disc particles themselves.

Radiation impinging on planetesimals generates both Poynting-Robertson drag and the Yarkovsky effect. The former is a long-standing topic of investigation of the evolution of white dwarf debris discs {\revv \citep{bocraf2011,rafikov2011a,rafikov2011b,metetal2012,rafgar2012,okuetal2023}}, but act non-negligibly only on objects which are smaller than the planetesimals which could undergo Type III migration. 

However, the Yarkovsky effect, when active, does have the potential to add or subtract from the Type III migration rate for the smallest migrating planetesimals: \cite{veras2020} showed that planetesimals with $R_{\rm pl} = 10^{-1}-10^1$~m can migrate on scales of $10^{-1}$~au over $10^4$~yr due to the Yarkovsky effect when the migration is consistently in the same direction. Nevertheless, for larger planetesimals, the effect is lessened because the migration timescale due to the Yarkovsky effect is directly proportional to $R_{\rm pl}$. 

{\rev However, there is a maximum body size below which the Yarkovsky effect is ineffectual; \cite{veretal2015a} estimated this value to be $10^{-1}-10^{1}$~m. Hence, the disc itself is not assumed to be altered significantly by the Yarkovsky effect \citep{rafikov2011b}. Nevertheless, the disc plays a role in the efficacy of the Yarkovsky effect because it may be limited due to absorption from disc material in which the planetesimal is embedded.}

Perhaps the most important effect of radiative drag is the effect on the (small) disc particles. This force drags particles to within the sublimation radius, creating a gaseous inner rim, and one whose surface density may differ from that in the remainder of the disc \citep{metetal2012,kenbro2017b}. Dust from the disc is hence continuously lost due to drag followed by sublimation, which might represent the primary factor in determining the disc lifetime.

\subsection{Gas drag and aeolian erosion}

Both \cite{ocolai2020} and \cite{malamudetal2021} have demonstrated the importance of gas drag in circularising the orbits of planetesimals embedded within white dwarf debris discs. However, because these discs contain gas, planet-disc migration would be negligible in these disc regions and hence not compete with migration induced by gas drag.  The situation is similar for aeolian erosion, which is a destructive mechanical process. \cite{rozetal2021} demonstrated that aeolian erosion would likely destroy planetesimals smaller than about 10~m in radius before much migration takes place, but only in the presence of gas.

\subsection{Ohmic heating, Alfv\'{e}n-wave drag and Lorentz drift}

A significant minority of detected white dwarfs have detectable magnetic fields: about 20 per cent {\revv have} fields stronger than 1~kG and 10 per cent with fields stronger than 1~MG \citep{feretal2015,holetal2015,lanbag2019}. These magnetic fields could drive the migration of conducting planetesimals through electromagnetic induction. Although this effect is known by different names -- Ohmic heating \citep{broken2019}, Lorentz drift \citep{verwol2019} and Alfv\'{e}n-wave drag \citep{zhaetal2021} -- the migration rates computed by all are similar and demonstrate that MG magnetic fields could migrate planetesimals at rates comparable to those induced by Type III migration. 

\subsection{Stellar oblateness and general relativity}

Both stellar oblateness and general relativity affect the evolution of orbiting bodies. However, the primary contribution of both effects is to alter an orbit's argument of pericentre, as opposed to its semimajor axis\footnote{Almost always these two effects are presented as averaged contributions, which average out to zero for the semimajor axis evolution. However, throughout the traversal of a single orbit, the effects do cause an orbit to drift. Nevertheless, for general relativity, the unaveraged effect is negligible \citep{veras2014}. For oblateness, we can compare to the Saturnian ring system, where the un-averaged drift for Saturn's rings is negligible. Saturn has an oblateness coefficient of $J_2 \sim 10^{-2}$. For a white dwarf, combining typical spin periods \citep{heretal2017} with quadrupole moment formulae \citep{sterne1939,waretal1976,spaetal2018} yields $J_2 \sim 10^{-9}$, which is seven orders of magnitude smaller.}. As a result, these effects do not compete with Type III-induced migration. Instead, these effects would help determine the collisional evolution within the disc, particularly for eccentric discs.

\subsection{Second-generation planetesimal formation}

Could a planetesimal which undergoes Type III migration have been formed in the disc itself? We think the answer is yes, but only for the most massive discs. \cite{vanetal2018} explored the prospects for ``second-generation formation" to take place, and found that the process can occur for $M_{\rm d} \gtrsim 10^{23}$kg discs which viscously spread outwards enough to allow particle coagulation to occur. {\rev Viscous outward spread beyond the Roche radii might be a common feature of these discs \citep{rafikov2011a,rafikov2011b}.}

As we have shown, such massive discs are particularly amenable to Type III migration, and this formation mechanism could provide a feasible initial condition for such a migrating planetesimal. Even if these planetesimals are formed exterior to the disc itself, eventually they will be dragged inward into the disc by one or more of the processes listed elsewhere in this Section.

\section{Summary}

The objective of this work was to explore whether any of the widely investigated six migration regimes in main-sequence planetary and satellite systems (Type I, Type II and Type III for gas-dominated and dust-dominated discs) are {\rev relevant on physically useful timescales} in white dwarf debris disc systems {\rev (Fig. \ref{DiscCartoon})}. 

We demonstrated throughout Section 3 and in Section 4.4 that five of these regimes are not relevant, with the lone exception being the dust-dominated Type III migration regime {\rev (Fig. \ref{Tab1})}. {\rev The primary reason for the irrelevance of most of the different migration regimes is because typical white dwarf disc masses are not sufficiently high, although several other parameters contribute to the migration timescales. Further, nonisothermal temperature profiles of discs introduce additional functional dependencies; we have also demonstrated that these cannot lower the migration timescales to relevant values.}

We then explored the nontrivial allowable parameter space in which dust-dominated Type III migration regime may take place (Figs. \ref{MdMpl}-\ref{Mplsigmai}) as well as the migration timescales themselves (Figs. \ref{DiscTraversal}-\ref{PatchTraversal}). {\rev Overall, we found that migration is possible in disc regions which are primarily more massive than about $10^{18}$~kg (Fig. \ref{MdMpl}) as long as the disc is sufficiently thin (Fig. \ref{Mplsigmai}), with few restrictions on disc particle size (Fig. \ref{MdRpart}) but more constraints on their eccentricities (Fig. \ref{Mdsigmae}).} Our results demonstrate that particulate Type III migration should be considered concurrently with the pantheon of other forces (Section 5) when modelling the evolution of planetesimals in dusty portions of white dwarf debris discs.



\section*{Acknowledgements}

{\revvv We thank Alexander J. Mustill and Judith Korth for thorough and astute comments which have improved the manuscript}.

\section*{Data Availability}

The numerical implementation of the equations presented in this paper are available upon reasonable request to the corresponding author.

\end{document}